\documentclass[12pt]{amsart}
\usepackage{geometry}                
\usepackage{graphicx}
\usepackage{amssymb}
\usepackage{epstopdf}
\usepackage{amsmath}
\usepackage{subcaption} 
\usepackage{booktabs} 
\usepackage{multirow} 
\usepackage{algorithm}
\usepackage{algpseudocode}

\usepackage{bm}
\usepackage{booktabs}
\usepackage[square,sort,comma,numbers]{natbib}
\newcommand{\nop}[1]{}
\usepackage{soul,color}
\usepackage{ragged2e}
\usepackage{booktabs,makecell,multirow,tabularx}
\usepackage{float}
\usepackage{hyperref}

\usepackage{enumitem}
\usepackage{makecell}

\DeclareMathOperator*{\argmax}{argmax}

\title[Estimating Bivariate Normal from Marginal Summaries]{Estimation of Bivariate Normal Distributions from Marginal Summaries in Clinical Trials}

\author[1]{Longwen Shang, Min Tsao, Xuekui Zhang*}

\begin{document}
\maketitle


\begin{center}
\textbf{ABSTRACT}
\end{center}
In certain privacy-sensitive scenarios within fields such as clinical trial simulations, federated learning, and distributed learning, researchers often face the challenge of estimating correlations between variables without access to individual-level data. To address this issue, we propose a novel method to estimate the correlation of bivariate normal variables using marginal information from multiple datasets. The method, based on maximum likelihood estimation (MLE), accommodates datasets with varying sample sizes and avoids reliance on sensitive information such as sample covariances, making it particularly suitable for privacy-restricted settings. Extensive simulation studies demonstrate the proposed method’s effectiveness in accurately estimating correlations and its robustness across diverse data configurations.

\section{Introduction}

Many statistical tasks from simulation and design planning to evidence synthesis and decision analysis require knowledge of the \emph{joint} distribution of two continuous variables, including their correlation coefficient \(\rho\).  In practice, however, analysts frequently possess only study-level \emph{marginal} summaries (sample means, variances and sizes) drawn from multiple independent data sets. Due to privacy laws \cite{GDPR_Art5, HIPAA_eCFR}, commercial sensitivities, or pragmatic constraints, the individual participant data (IPD) and within-study covariances are highly privacy‑sensitive \cite{Huth23Bioinf, NeurIPS22DPcov}. So the correlation coefficient \(\rho\) must be recovered from the marginal summaries alone.

The following are three real-world examples of the problem described above. \emph{(i) Clinical-trial simulation.} Sponsors planning Phase III programmes simulate efficacy–safety pairs (e.g., systolic‐pressure change versus liver-enzyme elevation).  Competitor trials of similar agents usually report per-arm means, standard deviations and sample sizes, but never release patient-level covariances.  Ignoring the correlation can distort power, safety-margin and futility‐stop calculations \cite{Leonov2020,Huang2022}, so it must be estimated from the marginal summaries. \emph{(ii) Cost-effectiveness analysis alongside trials.}  Health economic evaluations model the incremental joint distribution of costs and quality-adjusted life years (QALYs).  Published analyses typically include only marginal summaries, although guidelines recommend multivariate normal models that account for the cost–effect correlation \citep{WillanBriggs2006,Manca2005,Mitchell2016}. \emph{(iii) Diagnostic-test meta-analysis.}  Bivariate random-effects models that pool sensitivity and specificity require the within-study correlation, which is almost never reported; reviewers often resort to ad-hoc imputation \citep{Reitsma2005,Riley2007}.

There has been some work in the literature on estimating multivariate normal distributions when complete individual-level data are not available. Anderson (1957) proposed a method for estimating the parameters of multivariate normal distributions when some observations are missing \cite{Anderson1957}. Spruill and Gastwirth (1982) proposed estimators for the correlation coefficient based on grouped means and total variances, where groups are formed using order statistics of one variable, and only group-level summaries of the other are required \cite{Spruill1982}. More recent proposals combine a small set of IPD with external aggregates \citep{Raghunathan2003} or assume approximate covariances in multivariate meta-analysis \citep{LuAdes2004,Riley2009}. A domain-specific method also exists for binary outcomes \cite{Zhang2025}, but a general, privacy-preserving estimator for the joint distribution of two normal random variables based \emph{solely} on marginal summaries remains unavailable.

To address this challenge, we present a maximum-likelihood method for point and interval estimation of parameters of bivariate normal distributions that ({\emph a}) yields closed-form solutions for marginal means and variances, ({\emph b}) integrates out unobserved study-specific correlations, reducing estimation of \(\rho\) to a one-dimensional, box-constrained optimization, and ({\emph c}) employs an Olver asymptotic expansion to meet the challenge of maintaining numerical stability when computing the estimates for large sample sizes. Here,  ({\emph b}) and ({\emph c}) resolve two difficult issues unique to the implementation of the maximum-likelihood method for estimating bivariate normal distributions with marginal summaries. Simulation studies and a UK Biobank case-study show that the method is nearly unbiased, more efficient than naive or weighted-mean alternatives, and achieves nominal coverage with likelihood-ratio confidence intervals.  

The rest of this paper is organized as follows. Section \ref{section-method} introduces the theoretical framework of the proposed method. It also presents the optimization problem of computing the maximum likelihood estimates (MLE) of the distribution parameters. Section \ref{section-simulation} presents the results of simulation studies under various data simulation scenarios, including 1,000 experiment repetitions to assess its performance. In Section \ref{section-realdata}, the method is applied to real-world data set. Finally, Section \ref{section-conclusion} concludes with a discussion of the implications and future directions for this work.

\section{Method}
\label{section-method}

Estimating the joint distribution of two normal variables from marginal summaries is a challenging problem, particularly when only a single dataset is available, as it lacks sufficient information to estimate the covariance of the variables. However, when marginal summaries are available from multiple datasets drawn from the same population, they share a common underlying joint probability distribution function (PDF), so it becomes feasible to estimate the joint distribution. Below, we introduce the notation and formally define this problem.

\subsection{Problem Setup and Notation}

Let $ X $ and $ Y $ be two normally distributed variables:

$$
\begin{pmatrix} 
X \\ 
Y 
\end{pmatrix} 
\sim N \left( 
\boldsymbol{\mu}, 
\boldsymbol{\Sigma}
\right)
,
$$
where the mean vector is \(\boldsymbol{\mu}=
\begin{pmatrix} 
\mu_X \\ 
\mu_Y 
\end{pmatrix}\), with \(\mu_X, \mu_Y \in \mathbb{R}\), and the covariance matrix is
\(
\boldsymbol{\Sigma}=
\begin{pmatrix} 
\sigma_X^2 & \rho\,\sigma_X \sigma_Y \\ 
\rho\,\sigma_X \sigma_Y & \sigma_Y^2 
\end{pmatrix},
\)
where \(\sigma_X > 0\), \(\sigma_Y > 0\), and \(\rho \in (-1, 1)\) is the correlation between \(X\) and \(Y\). The full parameter vector is \(\theta = (\mu_X, \mu_Y, \sigma_X, \sigma_Y, \rho)\), and the parameter space is 
\(
\Theta = \mathbb{R}^2 \times (0, \infty)^2 \times (-1,1).
\)

Now, consider \( k \) independent datasets, all drawn from the same distribution above. For the \( i \)-th dataset (\( i = 1, \ldots, k \)), let \( n_i \) denote the sample size, 
$\overline{\mathbf{X}}_i=
\begin{pmatrix}
\bar{X}_i \\
\bar{Y}_i
\end{pmatrix}
= \frac{1}{n_i} \sum_{j=1}^{n_i}
\begin{pmatrix}
X_{ij} \\
Y_{ij}
\end{pmatrix}$
denotes the sample means, and 
$
\mathbf{S}_i =
\begin{pmatrix}
{{S_X}_i}^2 & {r_i} {{S_X}_i} {{S_Y}_i} \\
{r_i} {{S_X}_i} {{S_Y}_i} & {{S_Y}_i}^2
\end{pmatrix}
$
denotes the sample covariance matrix, where $r_i$ is the sample correlation coefficient. The sample means \(\bar{X}_i\), \(\bar{Y}_i\) and sample variances ${{S_X}_i}^2$, ${{S_Y}_i}^2$ are treated as random variables under the assumed model. Their observed values are denoted by \(\bar{x}_i, \bar{y}_i, {{s_X}_i}^2, {{s_Y}_i}^2\), respectively.

\vspace{1em}
\noindent
\textbf{Full Joint PDF.}
The sample means $\overline{\mathbf{X}}_i=
\begin{pmatrix}
\bar{X}_i \\
\bar{Y}_i
\end{pmatrix}$ follow a bivariate normal distribution

$$
\overline{\mathbf{X}}_i=
\begin{pmatrix}
\bar{X}_i \\
\bar{Y}_i
\end{pmatrix}
\sim N\left( 
\boldsymbol{\mu}, 
\frac{\boldsymbol{\Sigma}}{n_i} 
\right),
$$
where $n_i$ is the sample size of the \( i \)-th dataset. The probability density function is:

$$
f_{\overline{\mathbf{X}}_i}(\overline{\mathbf{x}}_i ; n_i) = 
\left[ \frac{1}{2\pi \sqrt{|\frac{1}{n_i} \boldsymbol{\Sigma}|}} \exp\left( -\frac{1}{2} 
(\overline{\mathbf{X}}_i - \boldsymbol{\mu})^T 
\left(\frac{\boldsymbol{\Sigma}}{n_i} \right)^{-1} 
(\overline{\mathbf{X}}_i - \boldsymbol{\mu})
\right) \right].
$$
The sample covariance matrix
$
\mathbf{S}_i =
\begin{pmatrix}
{{S_X}_i}^2 & {r_i} {{S_X}_i} {{S_Y}_i} \\
{r_i} {{S_X}_i} {{S_Y}_i} & {{S_Y}_i}^2
\end{pmatrix}
$
follows a Wishart distribution \cite{ref1} \cite{ref2}:
$$
\mathbf{S}_i \sim W_2(n_i-1, \frac{\boldsymbol{\Sigma}}{n_i-1}).
$$
The probability density function is:

$$
f_{\mathbf{S}_i}(\mathbf{s}_i ; n_i) = 
\frac
{|\mathbf{s}_i|^{(n_i-4)/2} \exp\left(-\frac{1}{2} \text{tr}[(\frac{\boldsymbol{\Sigma}}{n_i-1})^{-1} \mathbf{s}_i]\right)}
{2^{(n_i-1)} |\frac{\boldsymbol{\Sigma}}{n_i-1}|^{(n_i-1)/2} \Gamma_2\left(\frac{n_i-1}{2}\right)}
,
$$
where $s_i$ is positive definite.

For data from a multivariate normal distribution, the sample mean $\overline{\mathbf{X}}_i$ and sample covariance matrix $\mathbf{S}_i$ are independently distributed \cite{ref1}. So their full joint probability density function can be written as the product of their marginal density functions:

\begin{equation}
\label{full_joint_pdf}
f_{\overline{\mathbf{X}}_i, \mathbf{S}_i}(\overline{\mathbf{x}}_i, \mathbf{s}_i ; n_i) 
= f_{\overline{\mathbf{X}}_i}(\overline{\mathbf{x}}_i ; n_i) \cdot f_{\mathbf{S}_i}(\mathbf{s}_i ; n_i)
,
\end{equation}
where $s_i$ is positive definite. Specifically, this can be expressed as:

\begin{equation}
\label{full_joint_pdf_specific}
f_{\overline{X}_i,\overline{Y}_i, {S_X}_i, {S_Y}_i, r_i}(\overline{x}_i,\overline{y}_i, {s_X}_i, {s_Y}_i, r_i; n_i) = 
\frac{n_i}{2\pi \sigma_X \sigma_Y \sqrt{1 - \rho^2}}
\end{equation}

$$
\cdot
\exp\left[ -\frac{n_i}{2(1-\rho^2)} \left( \frac{(\bar{x}_i - \mu_X)^2}{\sigma_X^2} - \frac{2\rho (\bar{x}_i - \mu_X)(\bar{y}_i - \mu_Y)}{\sigma_X \sigma_Y} + \frac{(\bar{y}_i - \mu_Y)^2}{\sigma_Y^2} \right) \right]
$$

$$
\cdot \frac
{
[{{s_X}_i}^2 {{s_Y}_i}^2 (1 - {r_i}^2)]^{\frac{n_i-4}{2}} 
\exp\left[-\frac{n_i-1}{2(1 - \rho^2)} \left(\frac{{{s_X}_i}^2}{\sigma_X^2} + \frac{{{s_Y}_i}^2}{\sigma_Y^2} - 2 \frac{\rho {r_i}{{s_X}_i} {{s_Y}_i}}{\sigma_{X}\sigma_{Y}} \right)\right]
}
{
\left[\frac{2\sqrt{\sigma_X^2 \sigma_Y^2 (1 - \rho^2)}}{n_i-1}\right]^{n_i-1}
\Gamma_2\left(\frac{n_i-1}{2}\right)
}.
$$

\vspace{1em}
\noindent
\textbf{Problem Formulation.}
The primary objective of this study is to estimate the joint distribution of \(X\) and \(Y\), parameterized by \((\mu_X, \mu_Y, \sigma_X, \sigma_Y, \rho)\), using the observed summary statistics across \(k\) independent datasets:
\[
\{ (\bar{x}_i, \bar{y}_i, {s_{X}}_i^2, {s_{Y}}_i^2, n_i); \, i = 1, \ldots, k \}.
\]

Due to privacy or technical constraints, the individual-level data for each dataset are inaccessible, and only aggregated sample-level statistics are available. While the sample correlation coefficients \(r_i\) can theoretically provide crucial information about the dependency structure between \(X\) and \(Y\), these values are not observed directly and instead treated as latent variables. This introduces an additional layer of complexity to the estimation process.

The proposed approach involves maximizing the likelihood of the observed data while integrating over the unobserved \(r_i\) values, leveraging the information across multiple datasets to recover the joint PDF and provide robust parameter estimates.

\subsection{Maximum Likelihood Estimation (MLE) }

Since we have the full joint PDF (\ref{full_joint_pdf}) of the observed data \( \{(\bar{x}_i,\bar{y}_i,{s_X^2}_i,{s_Y^2}_i,n_i); i = 1, \ldots, k\} \), to estimate the parameters $(\mu_X,\mu_Y,\sigma_X,\sigma_Y,\rho)$ we use the method of maximum likelihood. Here, we present the key steps and formulas of maximum likelihood estimation. Detailed steps may be found in Appendix~\ref{appendix:mu MLE} and Appendix~\ref{appendix:sigma MLE}

\vspace{1em}
\noindent
\textbf{MLE for sample mean $\boldsymbol{\mu}$.}
The log-likelihood function for all sample means is:
$$
\ell(\boldsymbol{\mu} \mid \boldsymbol{\Sigma}) = \sum_{i=1}^k \log f_{\overline{\mathbf{X}}_i, \mathbf{S}_i}(\overline{\mathbf{x}}_i, \mathbf{s}_i, n_i; \boldsymbol{\mu}, \boldsymbol{\Sigma}).
$$

From the full joint PDF (\ref{full_joint_pdf}), taking the derivative of $\ell(\boldsymbol{\mu})$ with respect to $\boldsymbol{\mu}$ and setting it to zero, we obtain the MLE for $\boldsymbol{\mu}$
$$
\hat{\boldsymbol{\mu}} = \frac{\sum_{i=1}^k n_i \bar{\mathbf{x}}_i}{\sum_{i=1}^k n_i}.
$$

That is,

\begin{equation}
\label{mle_mu}
\hat{\mu}_X=\frac{\sum_{i=1}^k n_i \bar{x}_i}{\sum_{i=1}^k n_i}
, 
\hat{\mu}_Y=\frac{\sum_{i=1}^k n_i \bar{y}_i}{\sum_{i=1}^k n_i}.
\end{equation}

\vspace{1em}
\noindent
\textbf{MLE for sample covariance matrix $\boldsymbol{\Sigma}$.}
The log-likelihood function for the sample covariance matrix is:

\begin{equation}
\label{loglike_sigma}
\ell(\boldsymbol{\Sigma} \mid \boldsymbol{\mu}) = \sum_{i=1}^k \log f_{\overline{\mathbf{X}}_i, \mathbf{S}_i}(\overline{\mathbf{x}}_i, \mathbf{s}_i, n_i; \boldsymbol{\mu}, \boldsymbol{\Sigma}).
\end{equation}

From the full joint PDF \ref{full_joint_pdf}, taking the derivative of $\ell(\boldsymbol{\Sigma})$ with respect to $\boldsymbol{\Sigma}$ and setting it to zero, we have

$$\hat{\boldsymbol{\Sigma}} = 
\frac{\sum_{i=1}^{k} [ n_i (\overline{\mathbf{x}}_i - \boldsymbol{\mu})(\overline{\mathbf{x}}_i - \boldsymbol{\mu})^T + (n_i - 1) \mathbf{s}_i ]}{\sum_{i=1}^{k} n_i}$$

$$
= \begin{pmatrix}
\frac{\sum\limits_{i=1}^{k} \left[ n_i (\bar{x}_i - \mu_X)^2 + (n_i - 1) {{s_X}_i}^2 \right]}{\sum\limits_{i=1}^{k} n_i} & \frac{\sum\limits_{i=1}^{k} \left[ n_i (\bar{x}_i - \mu_X)(\bar{y}_i - \mu_Y) + (n_i - 1) {r_i} {{s_X}_i} {{s_Y}_i} \right]}{\sum\limits_{i=1}^{k} n_i} \\
\frac{\sum\limits_{i=1}^{k} \left[ n_i (\bar{x}_i - \mu_X)(\bar{y}_i - \mu_Y) + (n_i - 1) {r_i} {{s_X}_i} {{s_Y}_i} \right]}{\sum\limits_{i=1}^{k} n_i} & \frac{\sum\limits_{i=1}^{k} \left[ n_i (\bar{y}_i - \mu_Y)^2 + (n_i - 1) {{s_Y}_i}^2 \right]}{\sum\limits_{i=1}^{k} n_i}
\end{pmatrix}.
$$

Substituting the MLE's of $\mu_X$, $\mu_Y$ from equation \eqref{mle_mu} into the above estimation matrix and noting that the main diagonal of the estimation matrix does not contain the latent variable $r_i$, we obtain the MLE's for $\sigma_X$ and $\sigma_Y$

\begin{equation}
\label{mle_sigma}
\hat{\sigma}_X = \sqrt{\frac{\sum\limits_{i=1}^{k} \left[ n_i (\bar{x}_i - \hat{\mu}_X)^2 + (n_i - 1) {{s_X}_i}^2 \right]}{\sum\limits_{i=1}^{k} n_i}}
,
\hat{\sigma}_Y=\sqrt{\frac{\sum\limits_{i=1}^{k} \left[ n_i (\bar{y}_i - \hat{\mu}_Y)^2 + (n_i - 1) {{s_Y}_i}^2 \right]}{\sum\limits_{i=1}^{k} n_i}}.
\end{equation}

\vspace{1em}
\noindent
\textbf{MLE for the correlation coefficient $\rho$.}
The sample correlation $r_i$ for each study is not observed. To estimate the correlation $\rho$, we treat $r_i$ as a latent variable and integrate it out of the full joint likelihood. Following the integrated likelihood principle (Berger et al 1999~\cite{integratedlikelihood}), we have

\begin{equation}
\label{joint_pdf}
f_{\bar{X}_i, \bar{Y}_i, {S_X}_i, {S_Y}_i}(\bar{x}_i, \bar{y}_i, {s_X}_i, {s_Y}_i ; n_i) = 
\int_{-1}^{1} 
f_{\bar{X}_i,\bar{Y}_i, {S_X}_i, {S_Y}_i, r_i}(\bar{x}_i,\bar{y}_i, {s_X}_i, {s_Y}_i, r_i; n_i)
d{r_i}
\end{equation}

$$
= \frac{n_i}{2\pi \sigma_X \sigma_Y \sqrt{1 - \rho^2}}
\exp\left[ -\frac{n_i}{2(1-\rho^2)} \left( \frac{(\bar{x}_i - \mu_X)^2}{\sigma_X^2} - \frac{2\rho (\bar{x}_i - \mu_X)(\bar{y}_i - \mu_Y)}{\sigma_X \sigma_Y} + \frac{(\bar{y}_i - \mu_Y)^2}{\sigma_Y^2} \right) \right]
$$

$$
\cdot \frac
{
({{s_X}_i}^2 {{s_Y}_i}^2)^{\frac{n_i-4}{2}} 
\exp\left[-\frac{n_i-1}{2(1 - \rho^2)} \left(\frac{{{s_X}_i}^2}{\sigma_X^2} + \frac{{{s_Y}_i}^2}{\sigma_Y^2} \right)\right]
}
{
\left[\frac{2\sqrt{\sigma_X^2 \sigma_Y^2 (1 - \rho^2)}}{n_i-1}\right]^{n_i-1}
\Gamma_2\left(\frac{n_i-1}{2}\right)
}
\cdot \int_{-1}^{1} (1 - r_i^2)^{\frac{n_i - 4}{2}} \exp \left[ r_i \frac{(n_i - 1) \rho s_{X_i} s_{Y_i}}{(1 - \rho^2) \sigma_X \sigma_Y} \right]
d{r_i}.
$$

For

\begin{equation}
\label{oriint}
I_i = \int_{-1}^{1} (1 - r_i^2)^{a_i} e^{b_i r_i}
d{r_i}
\end{equation}

where $a_i = \frac{n_i - 4}{2}$ and $b_i = \frac{(n_i - 1) \rho s_{X_i} s_{Y_i}}{(1 - \rho^2) \sigma_X \sigma_Y}$. Let $r_i = \cos\theta, \quad \theta \in [0, \pi]$. Then $dr_i = -\sin\theta \, d\theta$. After simplification, we can get
$$I_i = \int_0^\pi e^{b_i \cos\theta} \sin^{2a_i+1} \theta d\theta.$$

The integral can be expressed using the modified Bessel function \cite{ref3}\cite{ref4}:
$$I_i = \sqrt{\pi} \left( \frac{2}{|b_i|} \right)^{a_i+\frac{1}{2}} \, \Gamma \left( a_i + 1 \right) I_{a_i+\frac{1}{2}} (|b_i|),$$
where \(\Gamma \left( a_i + 1 \right)\) is Gamma function and \(I_{a_i+\frac{1}{2}} (|b_i|)\) is the first kind modified Bessel function.


Incidentally, this class of integrals $I_i$ arises in statistical physics, where they describe orientation distributions of dipoles in external fields 
(see, e.g., Pathria and Beale \cite{pathriaStatisticalMechanics2011}). 
Their re-appearance in the present context of statistical estimation of bivariate normal distributions is intriguing. This connection to the distributions of dipoles leads to the following geometric interpretation of the integrals.
In the integral~(\ref{oriint}), the shape term \((1-r_i^{2})^{a_i}\) is proportional to the surface area element of a \((2a_i+3)\)-dimensional unit sphere at height \(r_i\). Hence larger sample sizes \(n_i\) (larger \(a_i\)) place more weight near \(r_i=0\), yielding a distribution with smaller variance when \(b_i=0\). Multiplying by the tilt term \(\exp(b_i r_i)\) exponentially favours positive poles if \(b_i>0\) and negative poles if \(b_i<0\); the magnitude \(|b_i|\) controls the strength of this alignment.  When \(|b_i|\) is small, the density remains almost symmetric; when \(|b_i|\) is large, most of the mass accumulates at \(r_i=\pm1\). In summary, sample size $n_i$ controls the shape term through $a_i=(n_i-4)/2$, while the tilt term 
$b_i=\dfrac{(n_i-1)\rho\,s_{X_i}s_{Y_i}}{(1-\rho^{2})\sigma_X\sigma_Y}$ 
takes its direction from $\operatorname{sign}(\rho)$ and its strength from $(n_i-1)\lvert\rho\rvert$.


We had derived, in (\ref{mle_mu}) and (\ref{mle_sigma}), closed-form solutions of the MLE's of $\mu_X$, $\mu_Y$, $\sigma_X$ and $\sigma_Y$ from the full joint PDF (\ref{full_joint_pdf}). However, under the integrated likelihood (\ref{joint_pdf}), closed form solutions are avaiable for the MLE's of $\mu_X$, $\mu_Y$, but not for $\sigma_X$, $\sigma_Y$, and $\rho$. The closed form solutions of MLE's for $\mu_X$ and $\mu_Y$ can be shown to coincide with those given by (\ref{mle_mu}). The MLE's for $\sigma_X$ and $\sigma_Y$ were found to be numerically nearly the same as the MLE's (\ref{mle_sigma}) under the full joint PDF (\ref{full_joint_pdf}). This is to be expected as the 
MLE's in (\ref{mle_sigma}) maximize the full likelihood regardless the underlying $r_i$, and as such they approximately maximize the integrated likelihood which may be viewed as a weighted version of the full likelihood over a range of $r_i$ values. For simplicity, without loss of numerical accuracy, we will treat (\ref{mle_sigma}) as the MLE's under the integrated likelihood. To estimate $\rho$, we substitute the MLE's of $\mu_X$, $\mu_Y$, $\sigma_X$ and $\sigma_Y$ from equations \eqref{mle_mu} and \eqref{mle_sigma} into the log-integrated-likelihood
$$
\ell(\rho \mid \mu_X, \mu_Y,\sigma_X, \sigma_Y) = \sum_{i=1}^k \log 
f_{\bar{X}_i, \bar{Y}_i, {S_X}_i, {S_Y}_i}(\bar{x}_i, \bar{y}_i, {s_X}_i, {s_Y}_i, n_i; \mu_X, \mu_Y,\sigma_X, \sigma_Y,\rho)
$$
to obtain the log-integrated-likelihood for $\rho$.
It follows that the MLE of $\rho$ is the maximizer of the log-integrated-likelihood subject to the constraint $-1\leq \rho \leq 1$,

\begin{align}
    \hat{\rho} = \argmax_{\rho \in [-1,1]} \sum_{i=1}^k \log f_{\bar{X}_i, \bar{Y}_i, {S_X}_i, {S_Y}_i}(\bar{x}_i, \bar{y}_i, {s_X}_i, {s_Y}_i, n_i; \hat{\mu}_X, \hat{\mu}_Y,\hat{\sigma}_X, \hat{\sigma}_Y,\rho).
\end{align}

The underlying boundary-constrained optimization problem can be solved by using the L-BFGS-B algorithm (Limited-memory Broyden-Fletcher-Goldfarb-Shanno with Box constraints) based R package \texttt{optim}. This algorithm builds on a quasi-Newton method that iteratively approaches the optimal solution using gradient information, while efficiently handling the parameter boundary constraints by storing only a limited amount of past update information. The initial value for L-BFGS-B algirthm can be set by a rough grid search in the interval $(-1,1)$.

\vspace{1em}
\noindent
\textbf{Asymptotic Approximation for Numerical Stability with Large $n_i$.}
For large values of $n_i$, directly evaluating the modified Bessel function can lead to numerical underflow, especially when computing $\log I_{a_i + \frac{1}{2}}(|b_i|)$. To address this, we adopt the following approximation:
\begin{equation}
\log I_{a_i+\frac{1}{2}}(|b_i|) 
\approx \kappa + \left( a_i + \tfrac{1}{2} \right) 
\log \left( \frac{|b_i|}{a_i + \tfrac{1}{2} + \kappa} \right) 
- \tfrac{1}{2} \log (2\pi \kappa),
\end{equation}
where $\kappa = \sqrt{b_i^2 + \left( a_i + \tfrac{1}{2} \right)^2}$. This is derived from the Olver asymptotic expansion for modified Bessel functions of large order, as documented in NIST Handbook of Mathematical Functions~\cite{NIST2010} and~\cite{olver1954bessel} by Olver (1954). This approximation is numerically stable and is used in our implementation when $\log I_{a_i+\frac{1}{2}}(|b_i|)$ returns infinity (e.g., when $n_i > 1000$). A detailed derivation of this expansion is provided in the Appendix~\ref{appendix:olver} and the empirical behavior and accuracy of this approximation under large-sample settings are evaluated in Section \ref{section-simulation}.

\subsection{Uncertainty Measure of Estimated Parameter $\rho$}

We focus on constructing confidence intervals (CIs) for the correlation coefficient $\rho$, which is typically the primary parameter of interest in bivariate analysis. Due to the structure of our estimation framework, we explore two approaches to interval estimation.

\vspace{1em}
\noindent
\textbf{The Normal Approximation Approach.}
A key advantage of the maximum likelihood approach to parameter estimation is that the MLE is in general consistent and asymptotically normally distributed. These properties make it possible to conduct hypothesis tests and construct confidence intervals (CI) for the unknown parameter, provided the standard error (SE) of the MLE can be effectively estimated. For the present problem, the 100\((1-\alpha)\)\% normal approximation based CI for $\rho$ is
\begin{align}
    CI_1({\rho}) &= \left( \hat{\rho} - z_{1 - \frac{\alpha}{2}} \cdot \hat{\textrm{SE}}(\hat{\rho}), \;\; \hat{\rho} + z_{1 - \frac{\alpha}{2}} \cdot \hat{\textrm{SE}}(\hat{\rho}) \right),
    \label{eq:CI1}
\end{align}
where $z_{1 - \frac{\alpha}{2}}$ is the $(1 - \frac{\alpha}{2})$th quantile of a standard normal distribution and $\hat{\textrm{SE}}(\hat{\rho})$ denotes the estimated SE of $\hat{\rho}$.

Due to the intractable closed-form expression of the Fisher information matrix caused by the presence of modified Bessel functions and Gamma functions in the marginal likelihood, we approximate the standard error of $\rho$ using the observed Fisher information, evaluated numerically via automatic differentiation of the log-likelihood function with respect to $\rho$. In our implementation, we used finite differences as provided in the \texttt{optim} routine in R, which computes the negative second derivative at the MLE. This numerical approach is justified as the log-likelihood is smooth and unimodal within the domain of \(\rho \in (-1, 1)\).

\vspace{1em}
\noindent
\textbf{The Likelihood Ratio Approach.}
The Gaussian approximation may not be ideal when the number of datasets, \( k \), is small. Additionally, it can result in confidence intervals (CIs) with bounds that fall outside the valid range of correlation, \([-1, 1]\). To address these issues, we propose an alternative approach to construct CIs for the estimated parameters using the likelihood ratio test.

The likelihood ratio test statistic for \( p_{11} \) is defined as:
\[
\text{LRTS}(\rho) = -2 \left[ \ell(\rho \mid \hat{\mu}_X, \hat{\mu}_Y, \hat{\sigma}_X, \hat{\sigma}_Y) - \ell(\hat{\rho} \mid \hat{\mu}_X, \hat{\mu}_Y, \hat{\sigma}_X, \hat{\sigma}_Y) \right],
\]
where \( \text{LRTS}(\rho) \) follows a Chi-squared distribution with one degree of freedom. Using this test statistic, the \((1 - \alpha)\)-level CI for \({\rho} \) is constructed as:
\begin{equation}
CI_2({\rho}) = \left\{ \rho : \text{LRTS}(\rho) \leq \chi^2_{1, \alpha} \right\}.
\label{eq:CI2}
\end{equation}

The bounds of this CI can be obtained by solving the equation \( \text{LRTS}(\rho) = \chi^2_{1, \alpha} \) for its two roots, where $\chi^2_{1, \alpha}$ represents the critical value of the Chi-squared distribution with one degree of freedom at significance level $\alpha$. 

\subsection{Large Sample Properties}
Let $\hat\theta=(\hat\mu_X,\hat\mu_Y,\hat\sigma_X,\hat\sigma_Y,\hat\rho)$ be the MLE of the true parameter vector $\theta_0$ obtained from $k$ independent studies. The following regularity conditions required by the asymptotic $M$‑estimation theory in Chapter 5 of van der Vaart (1998) \cite{vandervaart1998} are satisfied:
\begin{enumerate}[leftmargin=2em]
  \item[(i)] \textbf{Identifiability.} the map $(\mu_X,\mu_Y,\sigma_X,\sigma_Y,\rho)\mapsto f_\theta$ is injective;
  \item[(ii)] \textbf{Smoothness.} the Gaussian log‑likelihood is three‑times continuously differentiable on $$\Theta=\mathbb R^2\times(0,\infty)^2\times(-1,1),$$ and its derivatives are dominated by an integrable envelope;
  \item[(iii)] \textbf{i.i.d. sampling.} each study contributes independent observations from the same parametric law;
  \item[(iv)] \textbf{Information.} for $\sigma_X,\sigma_Y>0,\;|\rho|<1$, the Fisher information $I(\theta_0)$ is finite and non‑singular.
\end{enumerate}
Under conditions (i)–(iv), the estimators \(\hat{\theta}\) are consistent and asymptotically normal as \(k \to \infty\). Specifically,
\[
\hat{\theta} \xrightarrow{p} \theta_0, \quad \sqrt{k}(\hat{\theta} - \theta_0) \xrightarrow{d} \mathcal{N}(0, I^{-1}(\theta_0)),
\]
where \(I(\theta_0)\) is the Fisher information matrix. These properties have been empirically verified in an extensive simulation study. Section~\ref{section-simulation} contains the part of this simulation study concerning \(\hat{\rho}\) where it is shown that the bias of \(\hat{\rho}\) goes to zero and the empirical distribution \(\hat{\rho}\) approaches a normal distribution as \(k\) increases.


\section{Simulation Studies}
\label{section-simulation}

In this section, we evaluate the effectiveness of the proposed method by a comprehensive simulation study. For brevity, we focus on point and interval estimations for the correlation coefficient $\rho$. Similar results concerning the estimates of $\{\mu_X, \mu_Y, \sigma_X, \sigma_Y\}$ are relegated to the supplementary document.

\subsection{Simulation study design}

We consider 30 scenarios defined by different value combinations of the three factors: correlation coefficient $\rho$, sample sizes $n_i$, and the total number of studies (datasets) $k$. Without loss of generality, all samples are generated from a bivariate normal distribution with standardized marginals and correlation $\rho$, i.e, $
\mathcal{N} \left(
\begin{bmatrix}
0 \\
0
\end{bmatrix},
\begin{bmatrix}
1 & \rho\\
\rho & 1
\end{bmatrix}
\right)
$.

As shown in Table \ref{tab:simulation_settings}, the 30 scenarios include 3 levels of correlation strength: (i) $\rho=0.9$ representing strong correlation, (ii) $\rho=0.5$ representing moderate correlation, and (iii) $\rho=0.1$ representing weak correlation. More simulation results for correlation coefficient levels are provided in the supplementary document. For sample sizes of the datasets, we considered two cases with $n_i$ values randomly selected from (i) \( n_i \sim \text{DiscUnif}(100,200) \) representing small sample-size studies, and (ii) \( n_i \sim \text{DiscUnif}(800,1000) \) representing large-size studies. Here, DiscUnif(100, 200), for example, denotes the discrete uniform distribution supported on $\{ 100, 101, \dots, 200\}$. To explore the impact of the number of studies $k$, we
included five levels: $k=10, 20, 30, 40, 50$.


\begin{table}[ht]
    \centering
    \caption{Simulation scheme settings.}
    \label{tab:simulation_settings}
    \begin{tabular}{l l}
        \hline
        \textbf{Simulation Setting} & \textbf{Values} \\
        \hline
        Correlation (\(\rho\)) & Strong (0.9), Moderate (0.5), Weak (0.1) \\
        Number of Studies (\(k\)) & 10, 20, 30, 40, 50 \\
        Sample Sizes (\(n_i\)) & Small-size: \( n_i \sim \text{DiscUnif}(100,101,\dots 200) \) \\
                               & Large-size: \( n_i \sim \text{DiscUnif}(800,801,\dots 1000) \) \\
        \hline
    \end{tabular}
\end{table}

\subsection{Data Generation}

The summary-level datasets $\{(n_i, \bar{x}_i, \bar{y}_i, {s_{X}}_i^2, {s_{Y}}_i^2); \, i = 1, \ldots, k \}$ for simulation study are generated by Algorithm~\ref{alg:simulation}.

\begin{algorithm}
\caption{Simulating Summary-level Sample Data}
\label{alg:simulation}
\begin{algorithmic}[1]
\State \textbf{Input:} $k, n_{\min}, n_{\max}, \rho, \sigma_X, \sigma_Y, \mu_X, \mu_Y$.
\State Define covariance matrix:
\[
\Sigma =
\begin{bmatrix}
\sigma_X^2 & \rho \sigma_X \sigma_Y \\
\rho \sigma_X \sigma_Y & \sigma_Y^2
\end{bmatrix}
\]
\For{$i = 1$ to $k$}
    \State Randomly select $n_i$ from $\{ n_{\min}, n_{\min} + 1, \dots, n_{\max} \}$
    \State Sample $n_i$ points $(x_{ij}, y_{ij})$ from $\mathcal{N}((\mu_X, \mu_Y), \Sigma)$
    \State Compute sample mean:
    \begin{align*}
        \bar{x}_i &= \frac{1}{n_i} \sum_{j=1}^{n_i} x_{ij}, \quad
        \bar{y}_i = \frac{1}{n_i} \sum_{j=1}^{n_i} y_{ij}
    \end{align*}
    \State Compute sample variance:
    \begin{align*}
        {s^2_{X}}_i &= \frac{1}{n_i - 1} \sum_{j=1}^{n_i} (x_{ij} - \bar{x}_i)^2, \quad
        {s^2_{Y}}_i = \frac{1}{n_i - 1} \sum_{j=1}^{n_i} (y_{ij} - \bar{y}_i)^2
    \end{align*}
\EndFor
\State \textbf{Output:} $(n_i, \bar{x}_i, \bar{y}_i, {s^2_{X}}_i, {s^2_{Y}}_i)$ for $i = 1, \dots, k$.
\end{algorithmic}
\end{algorithm}

\subsection{Estimation Results and Evaluation}

For each scenario described in Table \ref{tab:simulation_settings}, we performed 1000 simulation runs, generating 1000 independent datasets using Algorithm~\ref{alg:simulation}. We computed 1000 independent estimates of $\rho$ and their estimated SEs with these datasets using the methodology outlined in Section \ref{section-method} and R package \texttt{optim}.

\begin{figure}[ht]
    \centering
    \includegraphics[width=\linewidth]{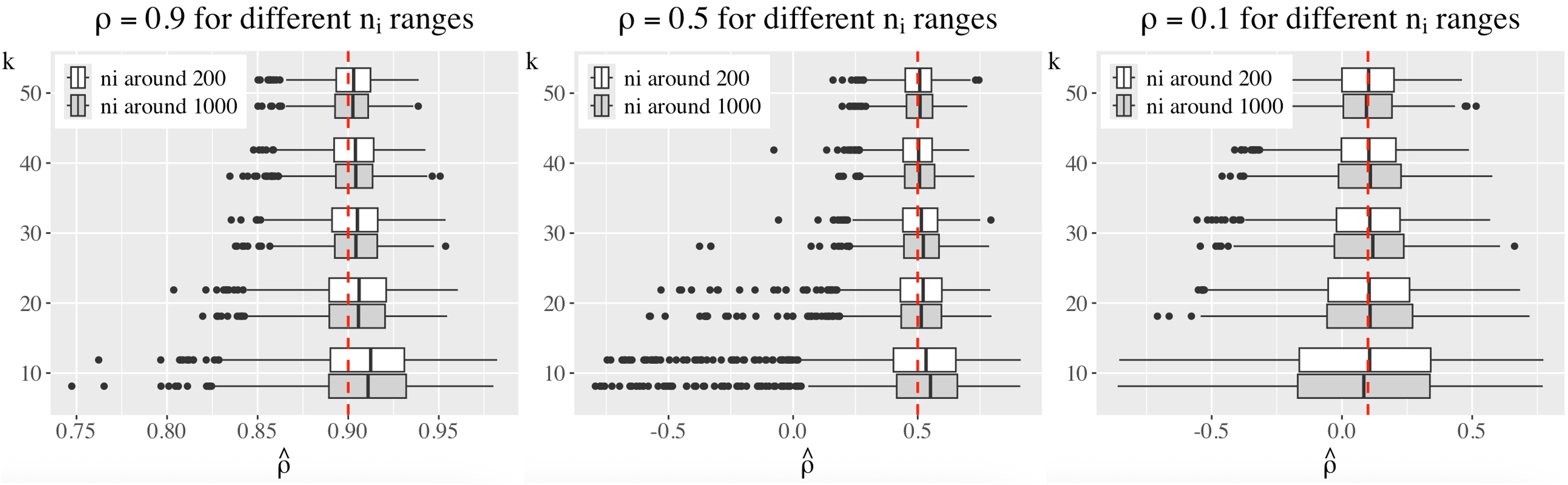}
    \caption{Boxplots of estimated $\rho$ values. Each boxplot is drawn with 1000 estimated values for a scenario. The red lines indicate the true $\rho$ values.}
    \label{est_rho_1}
\end{figure}

The 1000 estimates of $\rho$ for each scenario are summarized through a boxplot in Figure \ref{est_rho_1}. From this figure, we see that as $k$ increases, the median of the estimated results approaches the true value, and the estimation variance decreases. However, the impact of $n_i$ is small, reflecting the fact that summary-level data for each individual study does not contain information on $\rho$. This information is contained by between-study variations, so the larger the $k$, the more information about $\rho$ we have.

Table~\ref{tab:absmeanbias} numerically confirms the patterns observed in Figure~\ref{est_rho_1}. Specifically, the absolute estimated deviation of $\hat{\rho}$ consistently decreases as the number of studies $k$ increases across all correlation levels, providing quantitative support for the visual trend shown in the boxplots. This agreement between graphical and tabular results reinforces the conclusion that increasing the number of studies improves the accuracy of the estimated correlation, particularly when the true correlation is strong (e.g., $\rho = 0.9$). A table of signed bias values is provided in Appendix~\ref{appendix:bias}, confirming that the proposed estimator is nearly unbiased across all settings, particularly when correlation is strong.

\begin{table}[ht]
    \centering
    \caption{Estimated absolute deviation of $\hat{\rho}$ under different settings}
    \label{tab:absmeanbias}
    \begin{tabular}{c ccc ccc}
        \toprule
        & \multicolumn{3}{c}{$100 \leq n_i \leq 200$} & \multicolumn{3}{c}{$800 \leq n_i \leq 1000$} \\ 
        \cmidrule(lr){2-4} \cmidrule(lr){5-7}
        $k$ & $\rho = 0.9$ & $\rho = 0.5$ & $\rho = 0.1$
            & $\rho = 0.9$ & $\rho = 0.5$ & $\rho = 0.1$ \\
        \midrule
        10  &0.026  &0.180  &0.280  &0.027  &0.177  &0.282  \\
        20  &0.020  &0.108  &0.187  &0.018  &0.110  &0.185  \\
        30  &0.015  &0.083  &0.150  &0.015  &0.086  &0.158  \\
        40  &0.013  &0.069  &0.129  &0.013  &0.073  &0.137  \\
        50  &0.012  &0.064  &0.114  &0.011  &0.063  &0.110  \\
        \bottomrule
    \end{tabular}
\end{table}

To justify our proposed method of estimating the correlation using the joint PDF of sample means and sample variances, we need to compare it with the simple idea of using only means of the studies to estimate the correlation coefficient which we will refer to as the Naive Mean-Based Method. Details of this method may be found in Appendix~\ref{appendix:baseline1}. It is theoretically much simpler and if it is equally effective as our proposed method, then it would be the preferred method. A further refinement to this naive method is to incorporate sample sizes $n_i$. We call this refinement the Weighted Mean-Based Method. This method assigns higher weights to studies with larger sample sizes, but still does not account for sample variances; see Appendix~\ref{appendix:baseline2} for details. Both methods are computationally straightforward; however, they suffer from a critical drawback: they ignore sample variances, which may lead to the loss of valuable information. To demonstrate the effectiveness of our proposed method, we use these two approaches as baseline comparisons.

\begin{figure}[ht]
    \centering
    \includegraphics[width=\linewidth]{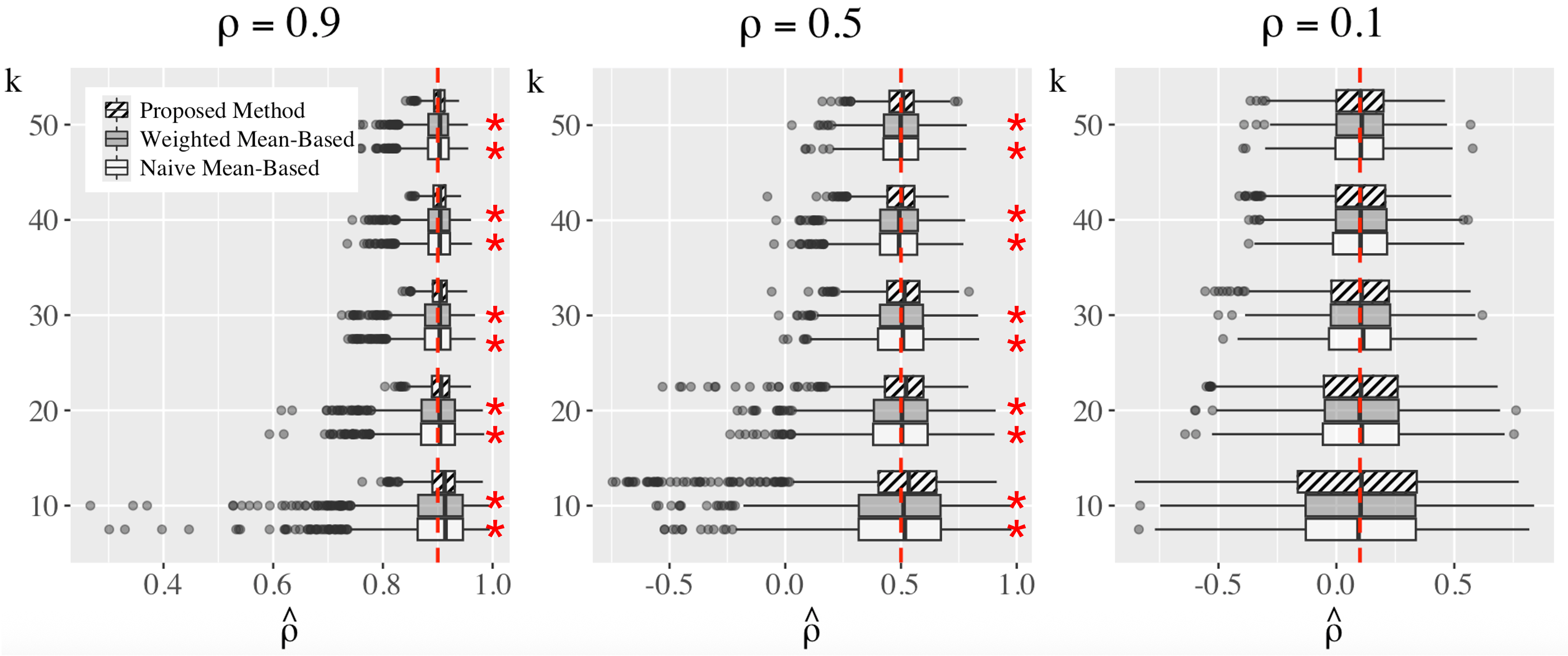}
    \caption{Boxplots of estimated $\rho$ values based on three different methods (low level $n_i$). The red lines indicate the true $\rho$ values. The asterisk (*) means our proposed method significantly outperforms a baseline method (\( p < 0.05 \)).}
    \label{est_rho_2}
\end{figure}

Since $n_i$ has a small impact on estimating $\rho$, we only retain scenarios with low level $n_i$ when comparing the baseline methods. Figure \ref{est_rho_2} presents the boxplots of 1,000 estimation results for the three methods under different settings. It can be observed that as the correlation strengthens, the estimates from our proposed method become increasingly accurate. When the correlation is low (e.g., $\rho=0.1$), there is no significant difference among the three methods.

\begin{figure}[ht]
    \centering
    \includegraphics[width=\linewidth]{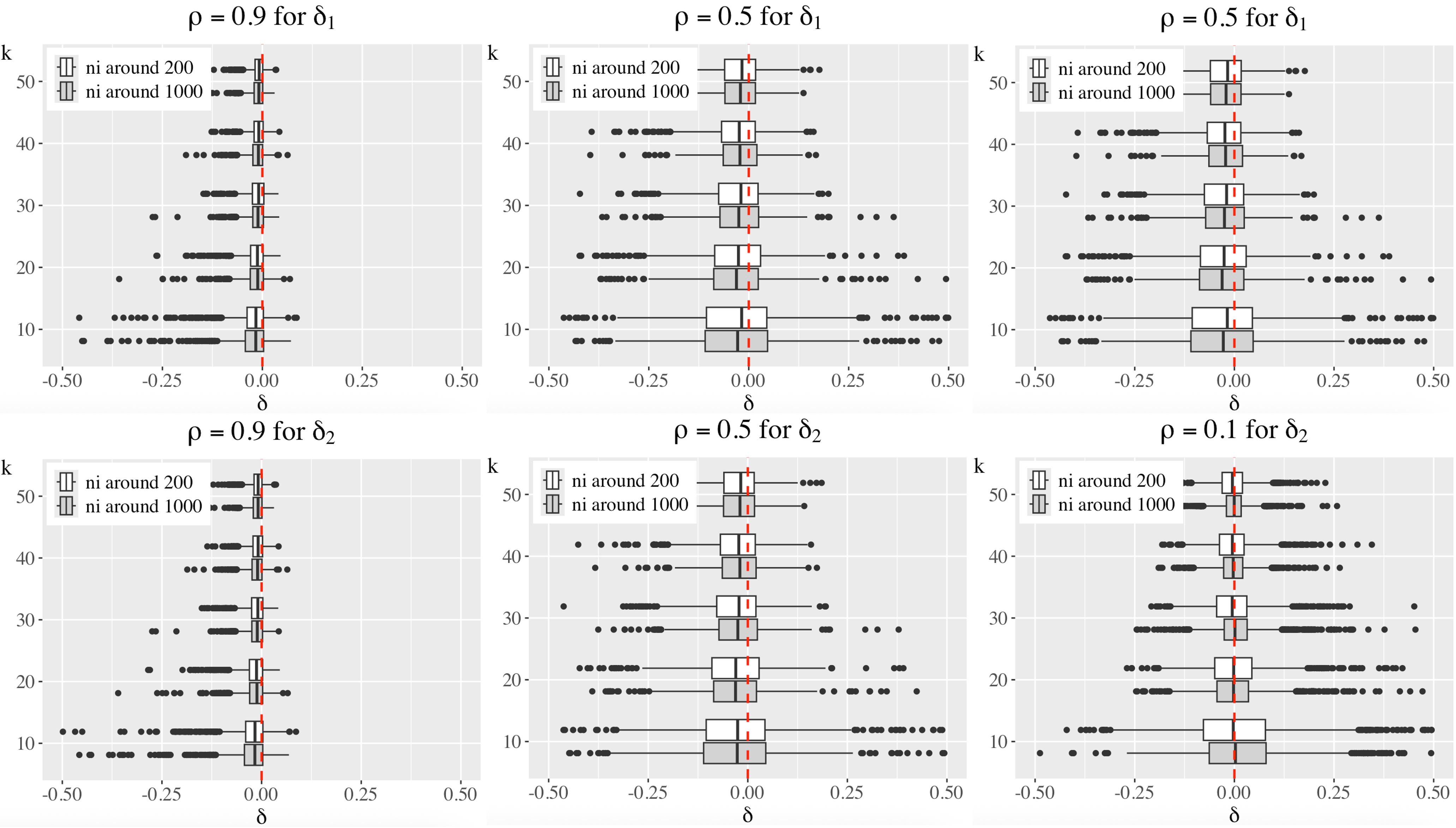}
    \caption{Boxplots of $\delta^{(j)} (j=1,2)$ values for two baseline methods. The red lines indicate 0.}
    \label{est_rho_delta}
\end{figure}

To quantitatively compare the estimation performance of our proposed method with the two baseline methods, we define the metric \( \delta_i^{(j)} \) as follows:

\begin{equation}
\delta^{(j)} = |\hat{\rho} - \rho_0| - |\tilde{\rho}_j - \rho_0|, \quad j = 1,2
\end{equation}
where \( \hat{\rho} \) represents the estimate obtained from our proposed method, \( \tilde{\rho}_j \) denotes the estimate from the \( j \)-th baseline method (\( j=1 \) for the Naive Mean-Based Method and \( j=2 \) for the Weighted Mean-Based Method), and \( \rho_0 \) is the true correlation coefficient. A negative \( \delta^{(j)} \) indicates that our method provides a more accurate estimate than the corresponding baseline method. We also conducted a one-sample t-test on \( \delta^{(j)} \) to determine whether our method provides significantly more accurate estimates. Specifically, we test the null hypothesis that the mean of \( \delta^{(j)} \) is zero against the alternative hypothesis that it is less than zero, indicating that our method tends to yield estimates closer to the true correlation coefficient compared to the baseline methods. In Figure~\ref{est_rho_2}, an asterisk (*) is added next to the boxplot of a baseline method when our estimator achieves a significantly smaller error than that baseline, as assessed by a one-sided, one-sample $t$-test of $\delta$ with null hypothesis $H_0: \mathbb{E}[\delta^{(j)}] = 0$, alternative $H_1: \mathbb{E}[\delta^{(j)}] < 0$, and significance level $\alpha = 0.05$ ($p < 0.05$). Figure \ref{est_rho_delta} presents the boxplots of \( \delta^{(1)} \) and \( \delta^{(2)} \). It can be observed that as the correlation strength increases, \( \delta \) tends to shift towards negative values, indicating that our proposed method provides more accurate correlation estimates when the true correlation is strong.

To assess the adequacy of the normal approximation to the finite sample distribution of $\hat{\rho}$, we generated the histogram and QQ plot for the estimates in each scenario. These plots (see Appendix~\ref{appendix:normal}) show that, except for cases where $k=10$, the normal approximation is adequate.

\begin{figure}[ht]
    \centering
    \includegraphics[width=\linewidth]{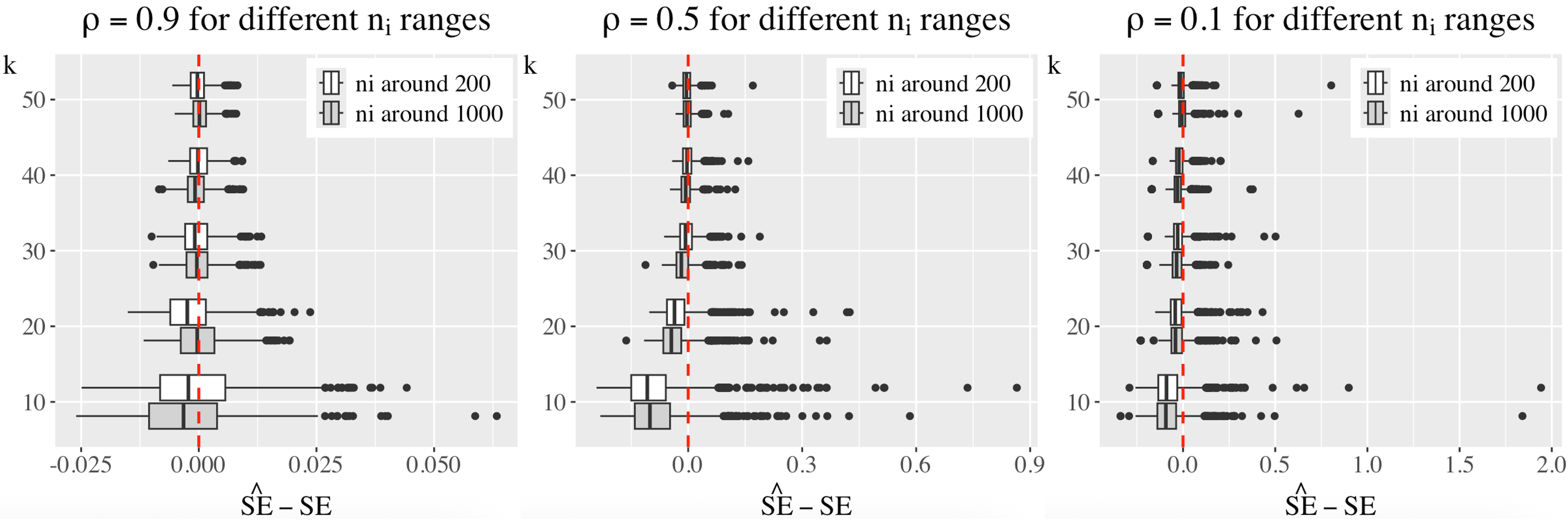}
    \caption{Boxplots of bias ($\hat{SE}$ $-$ true SE) of the estimated standard error for 30 scenarios. The red line represents $y=0$.}
    \label{est_SE}
\end{figure}

We also examined the estimated standard error $(\hat{SE})$ of $\hat{\rho}$ in each scenario, which was calculated using the observed Fisher information. Figure \ref{est_SE} shows the boxplot of the bias of the 1000 $\hat{SE}$ values. In order to compute these bias values, we needed the true SE of $\rho$ which is approximated by the sample standard deviation of the large sample of 1000 simulated $\hat{\rho}$ values in its place.

Finally, we computed confidence intervals for $\rho$ using both the normal approximation approach and the likelihood ratio approach. Table~\ref{tab:coverage_rates} and Table~\ref{tab:ave width} present the coverage rates and average widths of 95\% normal-based confidence intervals (\( CI_1 \)) and likelihood ratio based confidence intervals (\( CI_2 \)) under different settings over 1000 runs. Overall, both \(CI_1\) and \(CI_2\) exhibit some degree of under coverage, especially when \(k\) is small. However, \(CI_2\) consistently achieves coverage rates closer to the nominal 95\% level across all scenarios, indicating better calibration. As \(k\) increases, the coverage rates of both methods tend to improve and converge toward the nominal level, suggesting that larger meta-analytic sample sizes contribute to more reliable inference. The effect of within-study sample size \(n_i\) and the correlation level \(\rho\) on coverage is relatively modest. Regarding average widths, both \(CI_1\) and \(CI_2\) produce narrower intervals as \(k\) increases and as \(\rho\) becomes stronger. The two methods have comparable interval widths, but since the likelihood ratio confidence interval achieves better coverage while maintaining similar widths, it demonstrates superior efficiency.

\begin{table}[ht]
    \centering
    \caption{Coverage Rates of 95\% Confidence Intervals}
    \label{tab:coverage_rates}
    \begin{tabular}{cc ccc ccc ccc}
        \toprule
        & & \multicolumn{2}{c}{$\rho = 0.9$} & \multicolumn{2}{c}{$\rho = 0.5$} & \multicolumn{2}{c}{$\rho = 0.1$} \\ 
        \cmidrule(lr){3-4} \cmidrule(lr){5-6} \cmidrule(lr){7-8}
        $n_i$ size & $k$ & $CI_1$ & $CI_2$
           & $CI_1$ & $CI_2$
           & $CI_1$ & $CI_2$ \\
        \midrule
        \multirow{5}{*}{$100 \leq n_i \leq 200$}
        & 10  & 0.838 & 0.923 & 0.818 & 0.921 & 0.761 & 0.913 \\
        & 20  & 0.867 & 0.923 & 0.879 & 0.939 & 0.846 & 0.934 \\
        & 30  & 0.899 & 0.935 & 0.914 & 0.944 & 0.857 & 0.933 \\
        & 40  & 0.924 & 0.954 & 0.926 & 0.946 & 0.870 & 0.928 \\
        & 50  & 0.914 & 0.940 & 0.934 & 0.941 & 0.897 & 0.943 \\
        \midrule
        \multirow{5}{*}{$800 \leq n_i \leq 1000$} 
        & 10  & 0.827 & 0.910 & 0.805 & 0.913 & 0.753 & 0.927 \\
        & 20  & 0.905 & 0.947 & 0.885 & 0.940 & 0.858 & 0.946 \\
        & 30  & 0.907 & 0.948 & 0.915 & 0.947 & 0.837 & 0.933 \\
        & 40  & 0.917 & 0.938 & 0.906 & 0.935 & 0.859 & 0.927 \\
        & 50  & 0.931 & 0.954 & 0.925 & 0.944 & 0.886 & 0.957 \\
        \bottomrule
    \end{tabular}
\end{table}

\begin{table}[ht]
    \centering
    \caption{Average Widths of 95\% Confidence Intervals}
    \label{tab:ave width}
    \begin{tabular}{cc ccc ccc ccc}
        \toprule
        & & \multicolumn{2}{c}{$\rho = 0.9$} & \multicolumn{2}{c}{$\rho = 0.5$} & \multicolumn{2}{c}{$\rho = 0.1$} \\ 
        \cmidrule(lr){3-4} \cmidrule(lr){5-6} \cmidrule(lr){7-8}
        $n_i$ size & $k$ & $CI_1$ & $CI_2$
           & $CI_1$ & $CI_2$
           & $CI_1$ & $CI_2$ \\
        \midrule
        \multirow{5}{*}{$100 \leq n_i \leq 200$}
        & 10  & 0.116 & 0.139 & 0.697 & 0.773 & 1.019 & 0.944 \\
        & 20  & 0.087 & 0.094 & 0.494 & 0.536 & 0.789 & 0.732 \\
        & 30  & 0.071 & 0.075 & 0.394 & 0.416 & 0.658 & 0.623 \\
        & 40  & 0.062 & 0.064 & 0.346 & 0.362 & 0.576 & 0.548 \\
        & 50  & 0.056 & 0.057 & 0.305 & 0.314 & 0.525 & 0.500 \\
        \midrule
        \multirow{5}{*}{$800 \leq n_i \leq 1000$} 
        & 10  & 0.118 & 0.141 & 0.690 & 0.776 & 1.014 & 0.953 \\
        & 20  & 0.087 & 0.095 & 0.500 & 0.541 & 0.784 & 0.737 \\
        & 30  & 0.071 & 0.075 & 0.392 & 0.418 & 0.650 & 0.626 \\
        & 40  & 0.062 & 0.064 & 0.340 & 0.356 & 0.560 & 0.546 \\
        & 50  & 0.056 & 0.058 & 0.304 & 0.315 & 0.517 & 0.507 \\
        \bottomrule
    \end{tabular}
\end{table}

\subsection{Numerical Stability under Large Sample Sizes}

To further examine the robustness of our method under extreme sample sizes, we conducted an additional simulation where the sample sizes $n_i$ were independently drawn from a DiscUnif(2000, 5000) distribution. The number of studies was fixed at $k = 10, 20, 30, 40, 50$, and the true correlations were set to $\rho = 0.5$ and $\rho = 0.9$. The marginal means and variances were held constant at $\mu_X = \mu_Y = 0$ and $\sigma_X = \sigma_Y = 1$. In this setting, the direct evaluation of the modified Bessel function often results in numerical overflow or optimization failure. Therefore, we assess the numerical stability and estimation performance of our proposed asymptotic approximation when applied to such high $n_i$ datasets.

The results, included in Appendix~\ref{appendix:large ni}, show that the proposed method remains numerically stable and continues to yield accurate estimates, even when the Bessel function requires asymptotic approximation for computational feasibility. Compared to both the naive and weighted mean-based methods, our estimator exhibits lower variance and reduced bias across all values of $k$, confirming its robustness in large-sample regimes.


\section{Real Data Analysis}
\label{section-realdata}

To assess the practical utility of the proposed method in a real-world privacy-constrained scenario, we apply it to data from the UK Biobank dataset. The UK Biobank is a large-scale population-based cohort study comprising approximately 500,000 participants, recruited across the United Kingdom. The dataset spans multiple assessment centers, enabling the construction of pseudo-study-level data in a manner analogous to multicenter clinical trials. In this analysis, we focus on two continuous, approximately bivariate normal variables: diastolic blood pressure (automated reading, field 4079) and systolic blood pressure (automated reading, field 4080). In the overall UK Biobank population, the Pearson correlation between diastolic and systolic blood pressure is approximately 0.7.

To further examine the performance of the method under different total sample sizes and to reduce population heterogeneity across centers, we stratify the dataset by self-reported ethnic background (field 21000). Two subgroups are selected: the Irish population ($n = 12{,}109$) and the Chinese population ($n = 1{,}476$). These subgroups represent large and small-sample scenarios, respectively, and enable evaluation of the method under varying levels of data granularity that commonly arise in stratified or subgroup analyses. No additional inclusion or exclusion criteria were applied beyond self-reported ethnic background. In the Irish subgroup, the true correlation between diastolic and systolic blood pressure is 0.67, while in the Chinese subgroup it is 0.75. These subgroup-level ground truth values are computed from individual-level data and serve as benchmarks for evaluating the proposed method’s performance.

To simulate the summary-level data typically available in privacy-restricted settings, we treat each of the 22 UK Biobank assessment centers as an independent dataset, analogous to multi-center clinical trials. For each center, we compute only the marginal sample summaries: the sample means, variances, and sample sizes of the two variables, which serve as the observed inputs to our estimation procedure. The individual-level data, while not used in estimation, are retained to compute the true population parameters for evaluation. Incidentally, having the individual-level data and being able to compute the true population parameter for evaluating the accuracy of our method is a main reason for choosing the Biobank data.

\begin{figure}[ht]
    \centering
    \includegraphics[width=\linewidth]{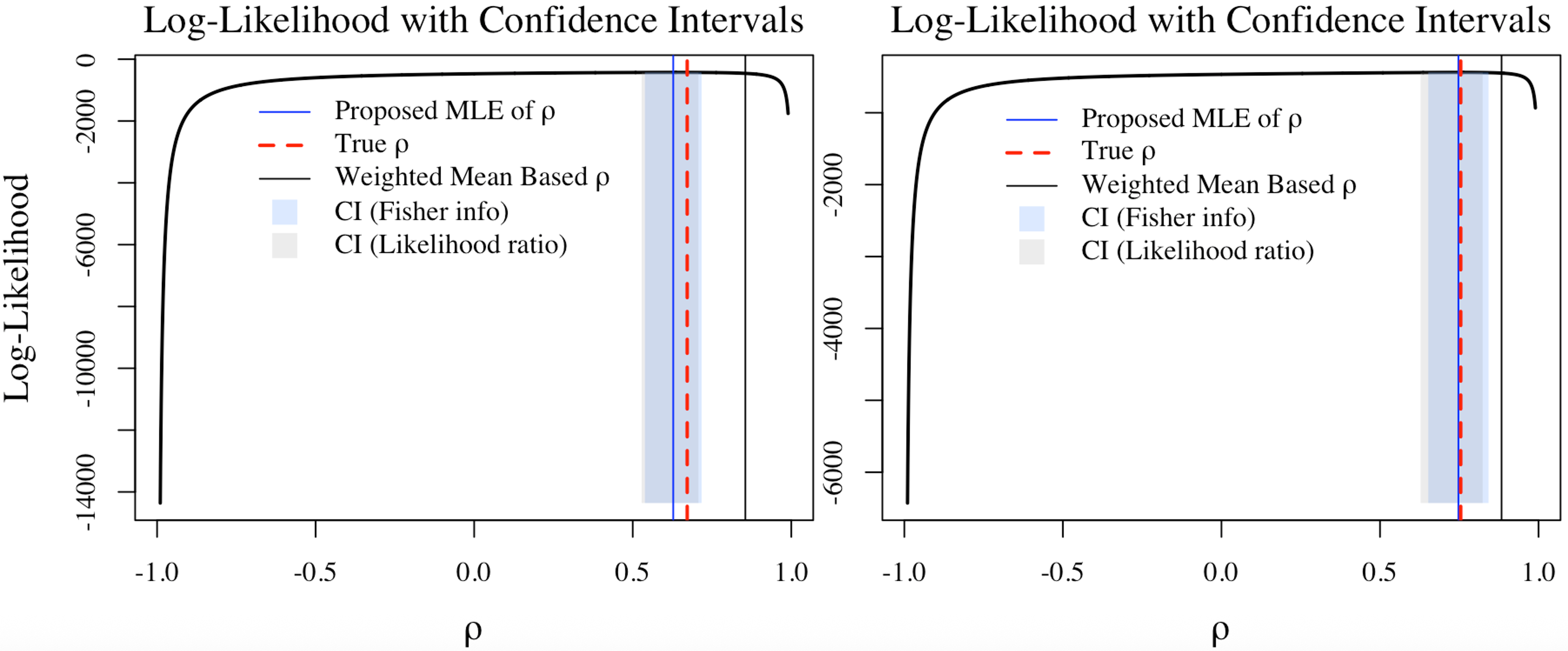}
    \caption{Point and interval estimation of the correlation coefficient $\rho$ between diastolic and systolic blood pressure. Results are shown separately for the Chinese (left) and Irish (right) subgroups.}
    \label{application_CI}
\end{figure}

For each subgroup, the proposed method is applied to the study-level summaries. As a benchmark, we compute the weighted mean-based estimator. Although a naive mean-based estimator was also considered, its performance closely mirrors that of the weighted version and is therefore omitted from further discussion. The estimated correlation coefficient based on our proposed method (proposed MLE of $\rho$) and that based on the weighted mean method are then compared to the ground truth. As shown in Fig. \ref{application_CI}, the proposed MLE is closer to the true value in both subgroups. The superior accuracy of the proposed MLE is particularly notable in the smaller Chinese subgroup where the weighted mean-based estimate has a large error.

This example demonstrates that the proposed method is capable of producing accurate estimates of the underlying correlation coefficients from summary-level data alone, even in scenarios with moderate to small per-center sample sizes. The results illustrate the method's applicability to federated or multi-center studies where individual-level data access is restricted.

\section{Discussion and Conclusion}
\label{section-conclusion}

The likelihood–based estimator developed here bridges an increasingly common gap between the rich multivariate models statisticians wish to fit and the privacy- or proprietary-driven data summaries they are actually given.  
Because it relies only on study-level means, variances and sample sizes, the method is immediately applicable to settings such as clinical-trial simulation, trial-based cost-effectiveness analysis, diagnostic-test meta-analysis and summary-statistic genomics.  
In a simulation study with a wide range of scenarios, it consistently outperformed the naive and weighted-mean alternatives, and produced likelihood-ratio intervals with near-nominal coverage even when the number of contributing studies was modest.

Extensive simulation experiments lead to four practical takeaways. First, increasing the number of studies $k$ gives more accurate estimates, whereas the within–study size $n_i$ has only a marginal impact; we recommend using at least $k\!\ge\!20$ whenever feasible. Second, the efficiency advantage of our method over the baselines grows with the true correlation~$\rho$. Third, estimated standard errors obtained from the observed Fisher information accurately track the empirical standard deviation, and both the normal-based and likelihood‑ratio confidence intervals work well; while the lengths of the two types of intervals are comparable, the observed coverage level of the likelihood‑ratio interval is consistently closer to the nominal level. Finally, the Olver asymptotic expansion used for the modified Bessel terms ensures numerical stability even for very large $n_i$ (e.g., $n_i\!>\!10^{3}$), preventing the overflow problem.

Although our motivating examples come from biostatistics, the framework is agnostic to scientific domain. One area of growing interest is federated analytics, where multiple data custodians perform local computations and share only aggregated statistics.  Our estimator could serve as a building block within such architectures—e.g.\ estimating cross-site correlations needed for pooled regression coefficients—without requiring additional protocol changes. At the same time, we have deliberately avoided tailoring the presentation to the mechanics of federated learning so as not to constrain readership to that community; the underlying inference problem predates and extends beyond any single computational paradigm.

Several extensions merit future investigation. First, the assumption of a common covariance matrix across studies could be relaxed by introducing random-effects or hierarchical priors, at the cost of one additional variance component.   Second, generalisation to higher-dimensional normals is algebraically straightforward but computationally demanding; sparsity-inducing penalties or composite-likelihood techniques may help. Third, coupling the estimator with formal privacy guarantees—differential-privacy noise at the study level or secure multicompartment computation across sites—would clarify its role in regulatory environments that specify explicit privacy budgets.

In conclusion, we have provided a numerically stable maximum-likelihood method for estimating the parameters of a bivariate normal distribution using only marginal summary statistics from multiple independent studies. By integrating out unobserved study-specific correlations, the method reduces the problem of estimating correlation to a one-dimensional optimization problem and yields closed-form solutions for all remaining parameters.   Extensive simulations and a real-data illustration demonstrated its accuracy and practical utility.   The estimator thus fills a methodological gap faced by analysts who must model joint behaviour under stringent data-sharing constraints, offering a principled alternative to ad-hoc imputation or the unrealistic assumption of independence.

\subsubsection*{Acknowledgments}
This work is supported by Canada Research Chair \#CRC-2021-00232 (XZ), and MSFHR Scholar Program SCH-2022-2553(XZ). This research was enabled in part by computational resource support provided by the Digital Research Alliance of Canada (https://alliancecan.ca) and Westgrid (https://www.westgrid.ca).

\subsubsection*{Competing interests}
The authors have no competing interests to declare. 

\subsubsection*{Author contributions statement}
 Study conceptualization and funding acquisition (XZ); supervision (XZ,MT);  methodology and experiment design (XZ,LS,MT); Computer experiment(LS); initial manuscript (LS). All authors contributed to the revision of the manuscript and have approved the final version.


\bibliographystyle{plain}
\bibliography{ref}

\newpage
\appendix
\section{Derivation of MLEs for $\mu_X$ and $\mu_Y$}
\label{appendix:mu MLE}

\vspace{1em}
\noindent
\textbf{Joint Probability Density Function.}
$$
f_{\overline{\mathbf{X}}_i, \mathbf{S}_i}(\overline{\mathbf{x}}_i, \mathbf{s}_i ; n_i) = 
\left[ \frac{1}{2\pi \sqrt{|\frac{1}{n_i} \boldsymbol{\Sigma}|}} \exp\left( -\frac{1}{2} 
(\overline{\mathbf{X}}_i - \boldsymbol{\mu})^T 
\left(\frac{1}{n_i} \boldsymbol{\Sigma}\right)^{-1} 
(\overline{\mathbf{X}}_i - \boldsymbol{\mu})
\right) \right]
$$
$$
\cdot \frac
{|\mathbf{s}_i|^{(n_i-4)/2} \exp\left(-\frac{1}{2} \text{tr}[(\frac{\boldsymbol{\Sigma}}{n_i-1})^{-1} \mathbf{s}_i]\right)}
{2^{(n_i-1)} |\frac{\boldsymbol{\Sigma}}{n_i-1}|^{(n_i-1)/2} \Gamma_2\left(\frac{n_i-1}{2}\right)},$$
where
$\boldsymbol{\mu}=
\begin{pmatrix}
\mu_{X}\\
\mu_{Y}
\end{pmatrix}
$,
$\boldsymbol{\Sigma}=
\begin{pmatrix} 
\sigma_X^2 & \rho\sigma_{X}\sigma_{Y} \\ 
\rho\sigma_{X}\sigma_{Y} & \sigma_Y^2 
\end{pmatrix}$,
$\overline{\mathbf{X}}_i=
\begin{pmatrix}
\bar{X}_i\\
\bar{Y}_i
\end{pmatrix} 
$
and
$
\mathbf{S}_i =
\begin{pmatrix}
{{S_X}_i}^2 & {r_i} {{S_X}_i} {{S_Y}_i} \\
{r_i} {{S_X}_i} {{S_Y}_i} & {{S_Y}_i}^2
\end{pmatrix}
$.

\vspace{1em}
\noindent
\textbf{Log-Likelihood Function.}
The likelihood function for all sample means is:
$$
L(\boldsymbol{\mu} \mid \boldsymbol{\Sigma}) = \prod_{i=1}^k f_{\overline{\mathbf{X}}_i, \mathbf{S}_i}(\overline{\mathbf{x}}_i, \mathbf{s}_i ; n_i).
$$
Taking the log of the likelihood function:
$$
\ell(\boldsymbol{\mu} \mid \boldsymbol{\Sigma}) = \sum_{i=1}^k \log f_{\overline{\mathbf{X}}_i, \mathbf{S}_i}(\overline{\mathbf{x}}_i, \mathbf{s}_i ; n_i).
$$
Ignoring the terms which are independent of $\boldsymbol{\mu}$:
$$
\ell(\boldsymbol{\mu} \mid \boldsymbol{\Sigma}) \propto -\frac{1}{2} \sum_{i=1}^k n_i (\bar{\mathbf{x}}_i - \boldsymbol{\mu})^T \boldsymbol{\Sigma}^{-1} (\bar{\mathbf{x}}_i - \boldsymbol{\mu}).
$$

\vspace{1em}
\noindent
\textbf{Derivative of the Log-Likelihood Function.}
Taking the derivative of $\ell(\boldsymbol{\mu} \mid \boldsymbol{\Sigma})$ with respect to $\boldsymbol{\mu}$ and setting it to zero:
$$
\frac{\partial \ell(\boldsymbol{\mu} \mid \boldsymbol{\Sigma})}{\partial \boldsymbol{\mu}} = -\frac{1}{2} \sum_{i=1}^k n_i \left( -2 \boldsymbol{\Sigma}^{-1} (\bar{\mathbf{x}}_i - \boldsymbol{\mu}) \right) = 0.
$$
Simplifying:
$$
\sum_{i=1}^k n_i \boldsymbol{\Sigma}^{-1} (\bar{\mathbf{x}}_i - \boldsymbol{\mu}) = 0.
$$
Then,
$$
\sum_{i=1}^k n_i \bar{\mathbf{x}}_i = \boldsymbol{\mu} \sum_{i=1}^k n_i.
$$
So,
$$
\hat{\boldsymbol{\mu}} = \frac{\sum_{i=1}^k n_i \bar{\mathbf{x}}_i}{\sum_{i=1}^k n_i}.
$$
That is,
$$
\hat{\mu}_X=\frac{\sum_{i=1}^k n_i \bar{x}_i}{\sum_{i=1}^k n_i}
, 
\hat{\mu}_Y=\frac{\sum_{i=1}^k n_i \bar{y}_i}{\sum_{i=1}^k n_i}.
$$

\section{Derivation of MLEs for $\sigma_X$ and $\sigma_Y$}
\label{appendix:sigma MLE}

\vspace{1em}
\noindent
\textbf{Joint Probability Density Function.}

$$
f_{\overline{\mathbf{X}}_i, \mathbf{S}_i}(\overline{\mathbf{x}}_i, \mathbf{s}_i ; n_i) = 
\left[ \frac{1}{2\pi \sqrt{|\frac{1}{n_i} \boldsymbol{\Sigma}|}} \exp\left( -\frac{1}{2} 
(\overline{\mathbf{X}}_i - \boldsymbol{\mu})^T 
\left(\frac{1}{n_i} \boldsymbol{\Sigma}\right)^{-1} 
(\overline{\mathbf{X}}_i - \boldsymbol{\mu})
\right) \right]
$$

$$
\cdot \frac
{|\mathbf{s}_i|^{(n_i-4)/2} \exp\left(-\frac{1}{2} \text{tr}[(\frac{\boldsymbol{\Sigma}}{n_i-1})^{-1} \mathbf{s}_i]\right)}
{2^{(n_i-1)} |\frac{\boldsymbol{\Sigma}}{n_i-1}|^{(n_i-1)/2} \Gamma_2\left(\frac{n_i-1}{2}\right)}.$$


\vspace{1em}
\noindent
\textbf{Log-Likelihood Function.}
The likelihood function for sample covariance matrix is:

$$
L(\boldsymbol{\Sigma} \mid \boldsymbol{\mu}) = \prod_{i=1}^k f_{\overline{\mathbf{X}}_i, \mathbf{S}_i}(\overline{\mathbf{x}}_i, \mathbf{s}_i ; n_i).
$$

Taking the log of the likelihood function:
$$
\ell(\boldsymbol{\Sigma} \mid \boldsymbol{\mu}) = \sum_{i=1}^k \log f_{\overline{\mathbf{X}}_i, \mathbf{S}_i}(\overline{\mathbf{x}}_i, \mathbf{s}_i ; n_i).
$$

Ignoring the terms which are independent of $\boldsymbol{\Sigma}$:
$$
\ell(\boldsymbol{\Sigma} \mid \boldsymbol{\mu}) \propto
- \frac{k}{2} \log \left|\boldsymbol{\Sigma}\right|
- \frac{1}{2} \sum_{i=1}^{k} n_i (\overline{\mathbf{X}}_i - \boldsymbol{\mu})^T \boldsymbol{\Sigma}^{-1} (\overline{\mathbf{X}}_i
- \boldsymbol{\mu}) - \frac{1}{2} \sum_{i=1}^{k} (n_i-1) \text{tr}\left[\boldsymbol{\Sigma}^{-1} \mathbf{s}_i\right]
- \frac{1}{2} \sum_{i=1}^{k} (n_i-1) \log \left|\boldsymbol{\Sigma}\right|
$$
$$
=
- \frac{1}{2} \sum_{i=1}^{k} n_i (\overline{\mathbf{X}}_i - \boldsymbol{\mu})^T \boldsymbol{\Sigma}^{-1} (\overline{\mathbf{X}}_i - \boldsymbol{\mu})
- \frac{1}{2} \sum_{i=1}^{k} (n_i-1) \text{tr}\left[\boldsymbol{\Sigma}^{-1} \mathbf{s}_i\right]
-\frac{1}{2} \sum_{i=1}^{k} n_i \log \left| \boldsymbol {\Sigma} \right|.
$$

\vspace{1em}
\noindent
\textbf{Derivative of the Log-Likelihood Function.}
Taking the derivative of $\ell(\boldsymbol{\Sigma} \mid \boldsymbol{\mu})$ with respect to $\boldsymbol{\Sigma}$ and setting it to zero:
$$
\frac{\partial \ell(\boldsymbol{\Sigma} \mid \boldsymbol{\mu})}{\partial \boldsymbol{\Sigma}} \propto
\frac{1}{2} \sum_{i=1}^{k} n_i \boldsymbol{\Sigma}^{-1} (\overline{\mathbf{X}}_i - \boldsymbol{\mu})(\overline{\mathbf{X}}_i - \boldsymbol{\mu})^T \boldsymbol{\Sigma}^{-1}
+ \frac{1}{2} \sum_{i=1}^{k} (n_i-1) \boldsymbol{\Sigma}^{-1} \mathbf{s}_i \boldsymbol{\Sigma}^{-1}
- \frac{1}{2} \sum_{i=1}^{k} n_i \boldsymbol{\Sigma}^{-1}
= 0.
$$

Simplifying:
$$
\sum_{i=1}^{k} n_i \boldsymbol{\Sigma}^{-1} (\overline{\mathbf{X}}_i - \boldsymbol{\mu})(\overline{\mathbf{X}}_i - \boldsymbol{\mu})^T \boldsymbol{\Sigma}^{-1}
+ \sum_{i=1}^{k} (n_i-1) \boldsymbol{\Sigma}^{-1} \mathbf{s}_i \boldsymbol{\Sigma}^{-1}
= \sum_{i=1}^{k} n_i \boldsymbol{\Sigma}^{-1}.
$$

Both sides of the equation are left-multiplied by a $\boldsymbol{\Sigma}^{-1}$:
$$
\sum_{i=1}^{k} n_i (\overline{\mathbf{X}}_i - \boldsymbol{\mu})(\overline{\mathbf{X}}_i - \boldsymbol{\mu})^T \boldsymbol{\Sigma}^{-1}
+ \sum_{i=1}^{k} (n_i-1) \mathbf{s}_i \boldsymbol{\Sigma}^{-1}
= \sum_{i=1}^{k} n_i \boldsymbol{I},
$$
$$
\left[
\sum_{i=1}^{k} n_i (\overline{\mathbf{X}}_i - \boldsymbol{\mu})(\overline{\mathbf{X}}_i - \boldsymbol{\mu})^T
+ \sum_{i=1}^{k} (n_i-1) \mathbf{s}_i
\right]
\boldsymbol{\Sigma}^{-1}
= \sum_{i=1}^{k} n_i \boldsymbol{I}.
$$
So,
$$\hat{\boldsymbol{\Sigma}} = 
\frac{\sum_{i=1}^{k} [ n_i (\overline{\mathbf{X}}_i - \boldsymbol{\mu})(\overline{\mathbf{X}}_i - \boldsymbol{\mu})^T + (n_i - 1) \mathbf{s}_i ]}{\sum_{i=1}^{k} n_i}.$$

That is,
$$
\hat{\boldsymbol{\Sigma}} = \begin{pmatrix}
\frac{\sum\limits_{i=1}^{k} \left[ n_i (\bar{X}_i - \mu_X)^2 + (n_i - 1) {{S_X}_i}^2 \right]}{\sum\limits_{i=1}^{k} n_i} & \frac{\sum\limits_{i=1}^{k} \left[ n_i (\bar{X}_i - \mu_X)(\bar{Y}_i - \mu_Y) + (n_i - 1) {r_i} {{S_X}_i} {{S_Y}_i} \right]}{\sum\limits_{i=1}^{k} n_i} \\
\frac{\sum\limits_{i=1}^{k} \left[ n_i (\bar{X}_i - \mu_X)(\bar{Y}_i - \mu_Y) + (n_i - 1) {r_i} {{S_X}_i} {{S_Y}_i} \right]}{\sum\limits_{i=1}^{k} n_i} & \frac{\sum\limits_{i=1}^{k} \left[ n_i (\bar{Y}_i - \mu_Y)^2 + (n_i - 1) {{S_Y}_i}^2 \right]}{\sum\limits_{i=1}^{k} n_i}
\end{pmatrix}.
$$
The estimated expression for the main diagonal does not contain the latent variable $r_i$. To estimate $\sigma_X$ and $\sigma_Y$, we substitute the MLE estimates of $\mu_X$, $\mu_Y$ and thus yields

$$
\hat{\sigma}_X = \sqrt{\frac{\sum\limits_{i=1}^{k} \left[ n_i (\bar{x}_i - \hat{\mu}_X)^2 + (n_i - 1) {{s_X}_i}^2 \right]}{\sum\limits_{i=1}^{k} n_i}}
,
\hat{\sigma}_Y=\sqrt{\frac{\sum\limits_{i=1}^{k} \left[ n_i (\bar{y}_i - \hat{\mu}_Y)^2 + (n_i - 1) {{s_Y}_i}^2 \right]}{\sum\limits_{i=1}^{k} n_i}}.
$$

\section{Baseline Methods}
\subsection{Naive Mean-Based Method}
\label{appendix:baseline1}
The naive mean-based method estimates the correlation coefficient \( \rho \) between two variables \( X \) and \( Y \) using only the sample means from \( k \) independent studies.

Let \( (\bar{x}_i, \bar{y}_i) \) denote the sample means of \( X \) and \( Y \) in the \( i \)-th study, for \( i = 1, 2, \dots, k \). Then, the naive estimator of the correlation coefficient is computed by directly applying the Pearson correlation formula to the $k$ observed sample means $(\bar{x}_i, \bar{y}_i)$:
\[
\hat{\rho}_{\text{naive}} = 
\frac{ \sum\limits_{i=1}^k (\bar{x}_i - \frac{1}{k} \sum\limits_{j=1}^k \bar{x}_j)(\bar{y}_i - \frac{1}{k} \sum\limits_{j=1}^k \bar{y}_j) }
     { \sqrt{ \sum\limits_{i=1}^k (\bar{x}_i - \frac{1}{k} \sum\limits_{j=1}^k \bar{x}_j)^2 } 
       \cdot \sqrt{ \sum\limits_{i=1}^k (\bar{y}_i - \frac{1}{k} \sum\limits_{j=1}^k \bar{y}_j)^2 } }.
\]

This method treats the study-level means as independent observations and computes their Pearson correlation directly. However, it does not account for sample sizes or within-study variances.

\subsection{Weighted Mean-Based Method}
\label{appendix:baseline2}
In our model, for each study $i = 1, \dots, k$, the observed sample mean vector 
$\overline{\mathbf{X}}_i = \begin{pmatrix} \bar{X}_i \\ \bar{Y}_i \end{pmatrix}$ 
follows a bivariate normal distribution:

\[
\overline{\mathbf{X}}_i \sim \mathcal{N}(\boldsymbol{\mu}, \frac{\boldsymbol{\Sigma}}{n_i}), \quad
\text{where} \quad 
\boldsymbol{\mu} = \begin{pmatrix} \mu_X \\ \mu_Y \end{pmatrix}, \quad
\boldsymbol{\Sigma} = \begin{pmatrix} \sigma_X^2 & \rho \sigma_X \sigma_Y \\ \rho \sigma_X \sigma_Y & \sigma_Y^2 \end{pmatrix}.
\]

That is,
$$
f_{\overline{\mathbf{X}}_i}(\overline{\mathbf{x}}_i; n_i) = 
\left[ \frac{1}{2\pi \sqrt{|\frac{1}{n_i} \boldsymbol{\Sigma}|}} \exp\left( -\frac{1}{2} 
(\overline{\mathbf{X}}_i - \boldsymbol{\mu})^T 
\left(\frac{1}{n_i} \boldsymbol{\Sigma}\right)^{-1} 
(\overline{\mathbf{X}}_i - \boldsymbol{\mu})
\right) \right].
$$

\vspace{1em}
\noindent
\textbf{Log-Likelihood Function.} Assuming independence across studies, the likelihood function is:
\[
L(\boldsymbol{\Sigma} \mid \boldsymbol{\mu}) = \prod_{i=1}^k f_{\overline{\mathbf{X}}_i}(\overline{\mathbf{x}}_i; n_i).
\]

Taking logs and dropping constant terms, we obtain:
\[
\ell(\boldsymbol{\Sigma} \mid \boldsymbol{\mu}) \propto 
- \frac{k}{2} \log |\boldsymbol{\Sigma}|
- \frac{1}{2} \sum_{i=1}^k n_i (\overline{\mathbf{X}}_i - \boldsymbol{\mu})^T 
\boldsymbol{\Sigma}^{-1} (\overline{\mathbf{X}}_i - \boldsymbol{\mu}).
\]

\vspace{1em}
\noindent
\textbf{Maximum Likelihood Estimator.} Taking the derivative with respect to $\boldsymbol{\Sigma}$ and setting to zero yields:
\[
\frac{\partial \ell(\boldsymbol{\Sigma} \mid \boldsymbol{\mu})}{\partial \boldsymbol{\Sigma}} \propto
- \frac{k}{2} \boldsymbol{\Sigma}^{-1}
+ \frac{1}{2} \sum_{i=1}^k n_i \boldsymbol{\Sigma}^{-1} 
(\overline{\mathbf{X}}_i - \boldsymbol{\mu})(\overline{\mathbf{X}}_i - \boldsymbol{\mu})^T 
\boldsymbol{\Sigma}^{-1} = 0.
\]

Simplifying:
$$
k\boldsymbol{\Sigma}^{-1}
=
\sum_{i=1}^{k} n_i \boldsymbol{\Sigma}^{-1} (\overline{\mathbf{X}}_i - \boldsymbol{\mu})(\overline{\mathbf{X}}_i - \boldsymbol{\mu})^T \boldsymbol{\Sigma}^{-1}.
$$

Multiplying both sides on the left by $\boldsymbol{\Sigma}$ and simplifying gives:
\[
k \mathbf{I} = \sum_{i=1}^k n_i (\overline{\mathbf{X}}_i - \boldsymbol{\mu})(\overline{\mathbf{X}}_i - \boldsymbol{\mu})^T \boldsymbol{\Sigma}^{-1},
\]
\[
\Rightarrow \hat{\boldsymbol{\Sigma}} = 
\frac{1}{k} \sum_{i=1}^k n_i (\overline{\mathbf{X}}_i - \boldsymbol{\mu})(\overline{\mathbf{X}}_i - \boldsymbol{\mu})^T.
\]

The element-wise form of $\hat{\boldsymbol{\Sigma}}$ is:
\[
\hat{\boldsymbol{\Sigma}} = 
\begin{pmatrix}
\frac{1}{k} \sum\limits_{i=1}^k n_i (\bar{X}_i - \mu_X)^2 
& 
\frac{1}{k} \sum\limits_{i=1}^k n_i (\bar{X}_i - \mu_X)(\bar{Y}_i - \mu_Y) 
\\
\frac{1}{k} \sum\limits_{i=1}^k n_i (\bar{X}_i - \mu_X)(\bar{Y}_i - \mu_Y) 
& 
\frac{1}{k} \sum\limits_{i=1}^k n_i (\bar{Y}_i - \mu_Y)^2 
\end{pmatrix}.
\]

Given the MLE of the covariance matrix $\hat{\boldsymbol{\Sigma}}$, we substitute the MLE estimates of $\mu_X$, $\mu_Y$ from equation \eqref{mle_mu} into the log-likelihood, and the correlation coefficient $\rho$ is estimated by:
\[
\hat{\rho} = 
\frac{ \sum\limits_{i=1}^k n_i (\bar{X}_i - \hat{\mu}_X)(\bar{Y}_i - \hat{\mu}_Y) }
     { \sqrt{ \sum\limits_{i=1}^k n_i (\bar{X}_i - \hat{\mu}_X)^2 } \cdot \sqrt{ \sum\limits_{i=1}^k n_i (\bar{Y}_i - \hat{\mu}_Y)^2 } }.
\]

The final estimator for $\rho$ obtained via the maximum likelihood approach has a form as a weighted Pearson correlation coefficient computed from the study-level means $(\bar{X}_i, \bar{Y}_i)$, where weights are proportional to the sample sizes $n_i$.

\section{Asymptotic Expansion for \texorpdfstring{$\log I_{a_i + \frac{1}{2}}(|b_i|)$}{log I\_a_i+1/2(|b_i|)}}
\label{appendix:olver}

\vspace{1em}
\noindent
\textbf{Modified Bessel Function of the First Kind.} The modified Bessel function of the first kind of order $\nu$ is defined as:
\[
I_\nu(x) = \sum_{k=0}^\infty \frac{1}{k! \, \Gamma(k + \nu + 1)} \left( \frac{x}{2} \right)^{2k + \nu}.
\]
It arises in numerous contexts including diffusion processes, random walks, and Bayesian models. However, when both $\nu$ and $x$ are large, evaluating $I_\nu(x)$ or its logarithm becomes numerically unstable.

\vspace{1em}
\noindent
\textbf{Olver’s Asymptotic Expansion (NIST §10.41(iii)).}
As $\nu \to \infty$ through positive real values, the uniform asymptotic expansion of $I_\nu(\nu z)$ for real $z > 0$ is:
\[
I_\nu(\nu z) \sim \frac{e^{\nu \eta}}{(2\pi \nu)^{1/2} (1 + z^2)^{1/4}} \sum_{k=0}^\infty \frac{U_k(p)}{\nu^k},
\]
where
\[
\eta = \sqrt{1 + z^2} + \log \left( \frac{z}{1 + \sqrt{1 + z^2}} \right), \quad p = (1 + z^2)^{-1/2},
\]
and $U_k(p)$ are polynomials of degree $3k$, with $U_0(p) = 1$.

\vspace{1em}
\noindent
\textbf{Logarithmic Form and Leading Term Approximation.} Taking the logarithm of the expansion and keeping only the leading term (since $\frac{U_k(p)}{\nu^k} = \mathcal{O}(1/\nu^k)$), we obtain:
\[
\log I_\nu(\nu z) \approx \nu \eta - \tfrac{1}{2} \log(2\pi \nu) - \tfrac{1}{4} \log(1 + z^2) + \log \left( 1 + \mathcal{O}\left( \tfrac{1}{\nu} \right) \right).
\]
Since $\log \left( 1 + \mathcal{O}(1/\nu) \right) = \mathcal{O}(1/\nu)$, this correction term can be safely ignored for large $\nu$ (e.g., $\nu \geq 1000$). To better understand the simplification, we start by expressing the asymptotic expansion in terms of $x = \nu z$ and recall that:
\[
\eta = \sqrt{1 + z^2} + \log \left( \frac{z}{1 + \sqrt{1 + z^2}} \right).
\]

Multiplying by $\nu$, we obtain:
\begin{align*}
\nu \eta 
&= \nu \sqrt{1 + z^2} + \nu \log \left( \frac{z}{1 + \sqrt{1 + z^2}} \right) \\
&= \kappa + \nu \log \left( \frac{z}{1 + \sqrt{1 + z^2}} \right), \quad \text{where } \kappa := \nu \sqrt{1 + z^2} = \sqrt{x^2 + \nu^2}.
\end{align*}

Now observe that:
\[
\frac{z}{1 + \sqrt{1 + z^2}} = \frac{x / \nu}{1 + \kappa / \nu} = \frac{x}{\nu + \kappa}.
\]

Therefore,
\[
\nu \log \left( \frac{z}{1 + \sqrt{1 + z^2}} \right) = \nu \log \left( \frac{x}{\nu + \kappa} \right),
\]
\[
\nu \eta = \kappa + \nu \log \left( \frac{x}{\nu + \kappa} \right).
\]

Now, consider the remaining log terms:
\[
- \tfrac{1}{2} \log(2\pi \nu) - \tfrac{1}{4} \log(1 + z^2)
= - \tfrac{1}{2} \log(2\pi \nu \sqrt{1 + z^2}).
\]

Note that $\nu \sqrt{1 + z^2} = \kappa$, so we substitute:
\[
- \tfrac{1}{2} \log(2\pi \nu \sqrt{1 + z^2}) = - \tfrac{1}{2} \log(2\pi \kappa).
\]

Putting everything together, we get the simplified expression:
\[
\log I_\nu(x) \approx \kappa + \nu \log \left( \frac{x}{\nu + \kappa} \right) - \tfrac{1}{2} \log(2\pi \kappa), \quad \text{where } \kappa = \sqrt{x^2 + \nu^2}.
\]

In our context, we use:
\[
\nu = a_i + \tfrac{1}{2} = \tfrac{n_i - 3}{2}, \quad x = |b_i| = \left| \frac{(n_i - 1)\rho s_{Xi}s_{Yi}}{(1 - \rho^2) \sigma_X \sigma_Y} \right|.
\]
Substituting into the formula gives the following numerically stable approximation used in our method:
\[
\log I_{a_i + \frac{1}{2}}(|b_i|) 
\approx \kappa + \left( a_i + \tfrac{1}{2} \right) 
\log \left( \frac{|b_i|}{a_i + \tfrac{1}{2} + \kappa} \right) 
- \tfrac{1}{2} \log (2\pi \kappa),
\quad \text{where } \kappa = \sqrt{b_i^2 + \left( a_i + \tfrac{1}{2} \right)^2}.
\]

This expression is used in our MLE implementation whenever the direct evaluation of $\log I_{a_i + \frac{1}{2}}(|b_i|)$ becomes numerically unstable (e.g., results in infty when $n_i$ is large).

\section{Additional Simulation Results}

\subsection{Bias of the estimated $\hat{\rho}$}
\label{appendix:bias}
Table~\ref{tab:meanbias} reports the signed bias of $\hat{\rho}$ across different simulation settings. The overall bias is small, particularly when the true correlation is strong ($\rho = 0.9$), indicating that the estimator is nearly unbiased. In this high correlation scenario, the bias decreases noticeably as the number of studies $k$ increases, consistent with asymptotic unbiasedness. For moderate and weak correlations ($\rho = 0.5$ and $\rho = 0.1$), the decreasing trend is less pronounced. This may be attributed to numerical inaccuracies introduced during the optimization and Bessel function approximation steps, which can obscure the convergence pattern when $k$ becomes large. However, the bias remains very small. If we define the relative bias as
\[
\text{Relative Bias} = \frac{\text{Bias}}{\rho_{true}},
\]
we find that for $k > 10$, the relative bias lies between 0.1\% and 8\% across all settings, supporting the conclusion that the proposed estimator is nearly unbiased in practice.

\begin{table}[ht]
    \centering
    \caption{Estimated bias of $\hat{\rho}$ under different settings}
    \label{tab:meanbias}
    \begin{tabular}{c ccc ccc}
        \toprule
        & \multicolumn{3}{c}{$100 \leq n_i \leq 200$} & \multicolumn{3}{c}{$800 \leq n_i \leq 1000$} \\ 
        \cmidrule(lr){2-4} \cmidrule(lr){5-7}
        $k$ & $\rho = 0.9$ & $\rho = 0.5$ & $\rho = 0.1$
            & $\rho = 0.9$ & $\rho = 0.5$ & $\rho = 0.1$ \\
        \midrule
        10  &0.0094  &-0.0183  &-0.0185  &0.0083  &-0.0043  &-0.0290  \\
        20  &0.0039  &0.0018  &-0.0075  &0.0039  &-0.0066  &-0.0069  \\
        30  &0.0036  &0.0062  &-0.0082  &0.0035  &0.0081  &0.0003  \\
        40  &0.0029  &-0.0021  &-0.0040  &0.0026  &0.0044  &-0.0019  \\
        50  &0.0022  &0.0014  &-0.0030  &0.0016  &0.0050  &-0.0055  \\
        \bottomrule
    \end{tabular}
\end{table}

\subsection{Normality Assessment of $\rho$ Estimation Results}
\label{appendix:normal}
\begin{figure}[ht]
    \centering
    \includegraphics[width=\linewidth]{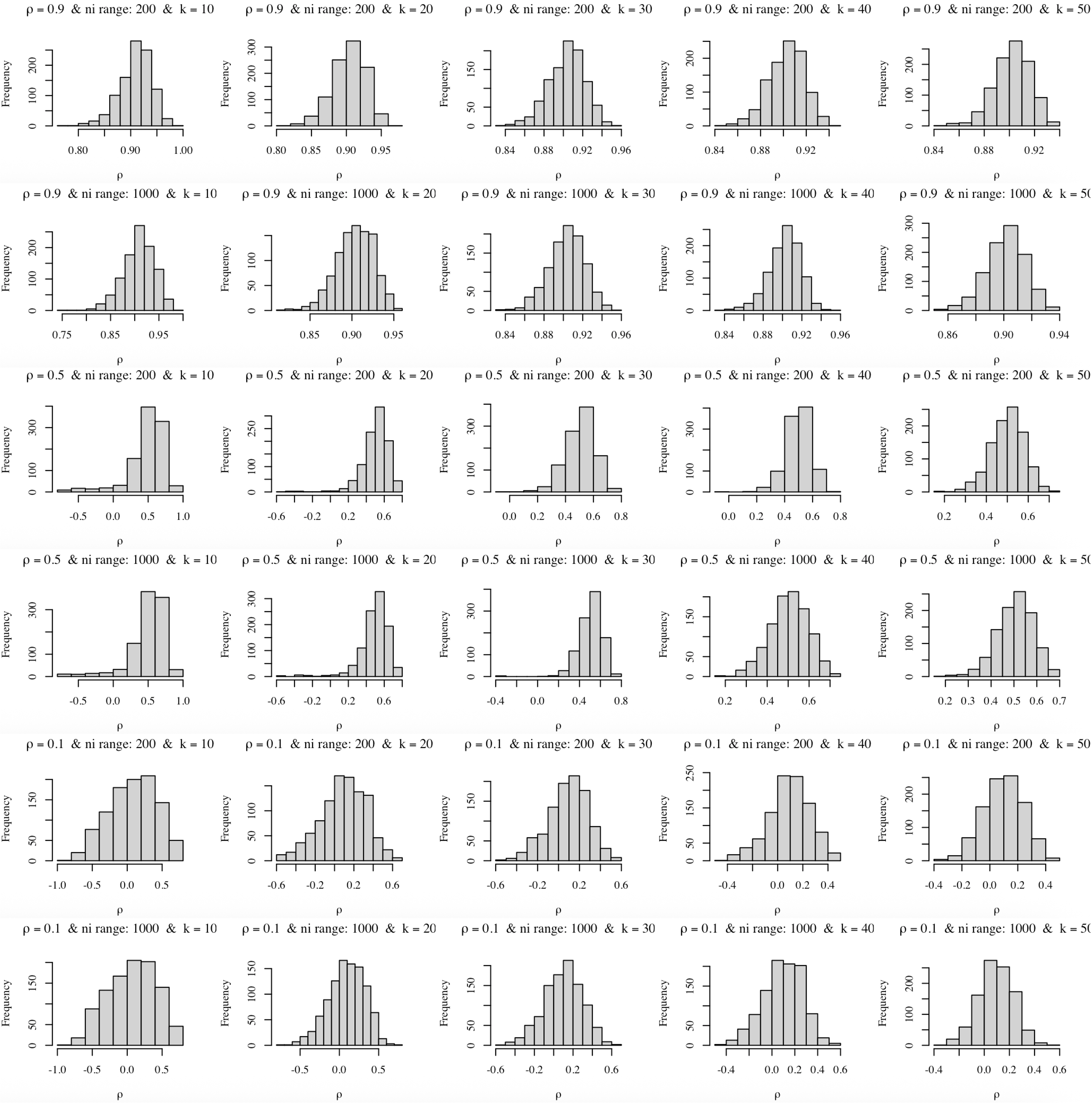}
    \caption{Histograms of the estimated \(\rho\) obtained from 1000 repetitions of the experiment conducted under 30 different settings. Each histogram corresponds to a specific experimental configuration, illustrating the distribution of \(\rho\) estimates.}
    \label{hist}
\end{figure}

\begin{figure}[ht]
    \centering
    \includegraphics[width=0.8\linewidth]{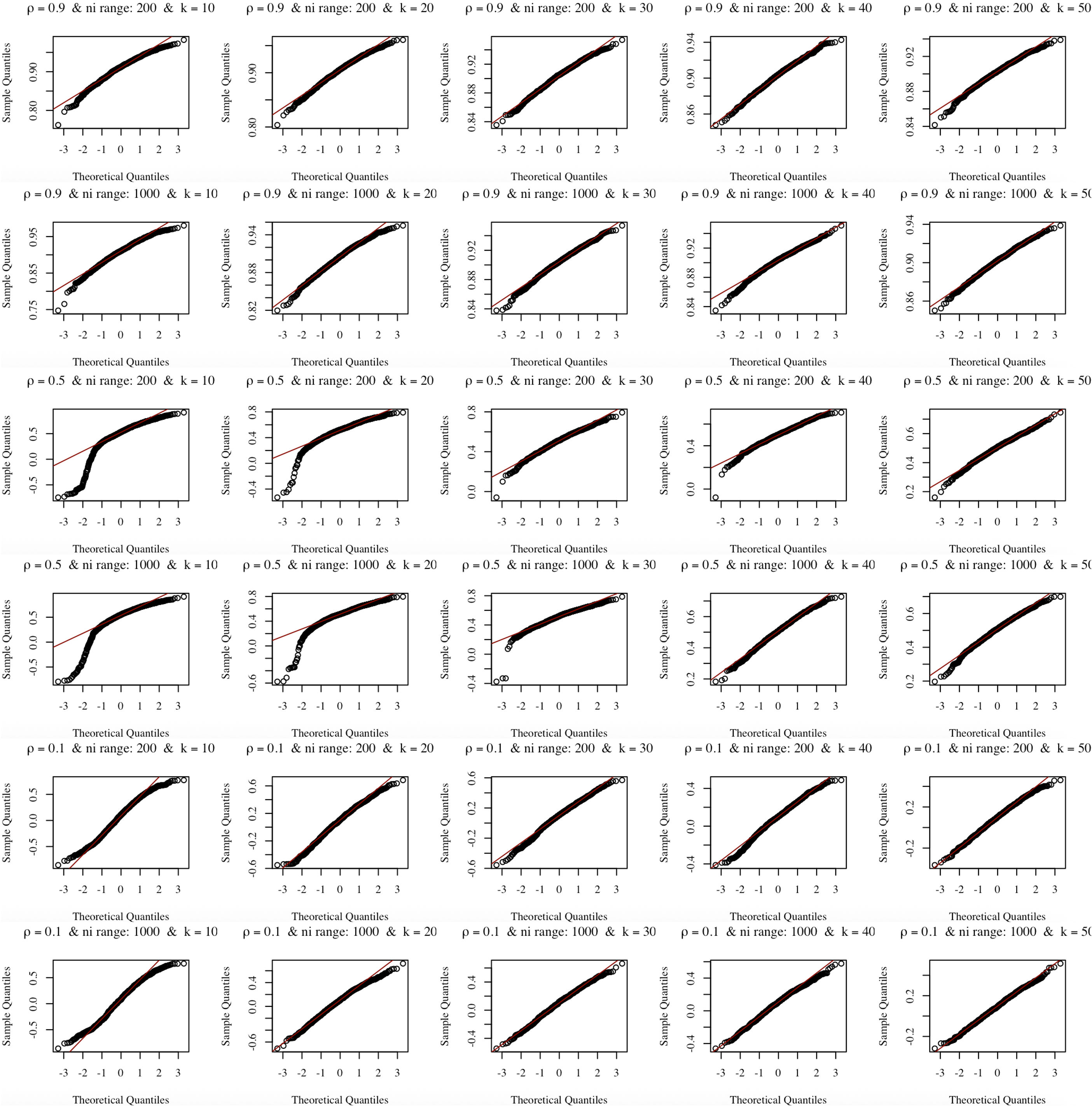}
    \caption{QQ plots comparing the quantiles of the estimated \(\rho\) distributions to the theoretical quantiles of a standard normal distribution. These plots validate the approximate normality of the \(\rho\) estimates, except in cases with small sample sizes (e.g., \(k=10\)).}
    \label{qqplots}
\end{figure}

\subsection{Numerical Stability under Large Sample Sizes}
\label{appendix:large ni}

\begin{figure}[ht]
    \centering
    \includegraphics[width= \textwidth]{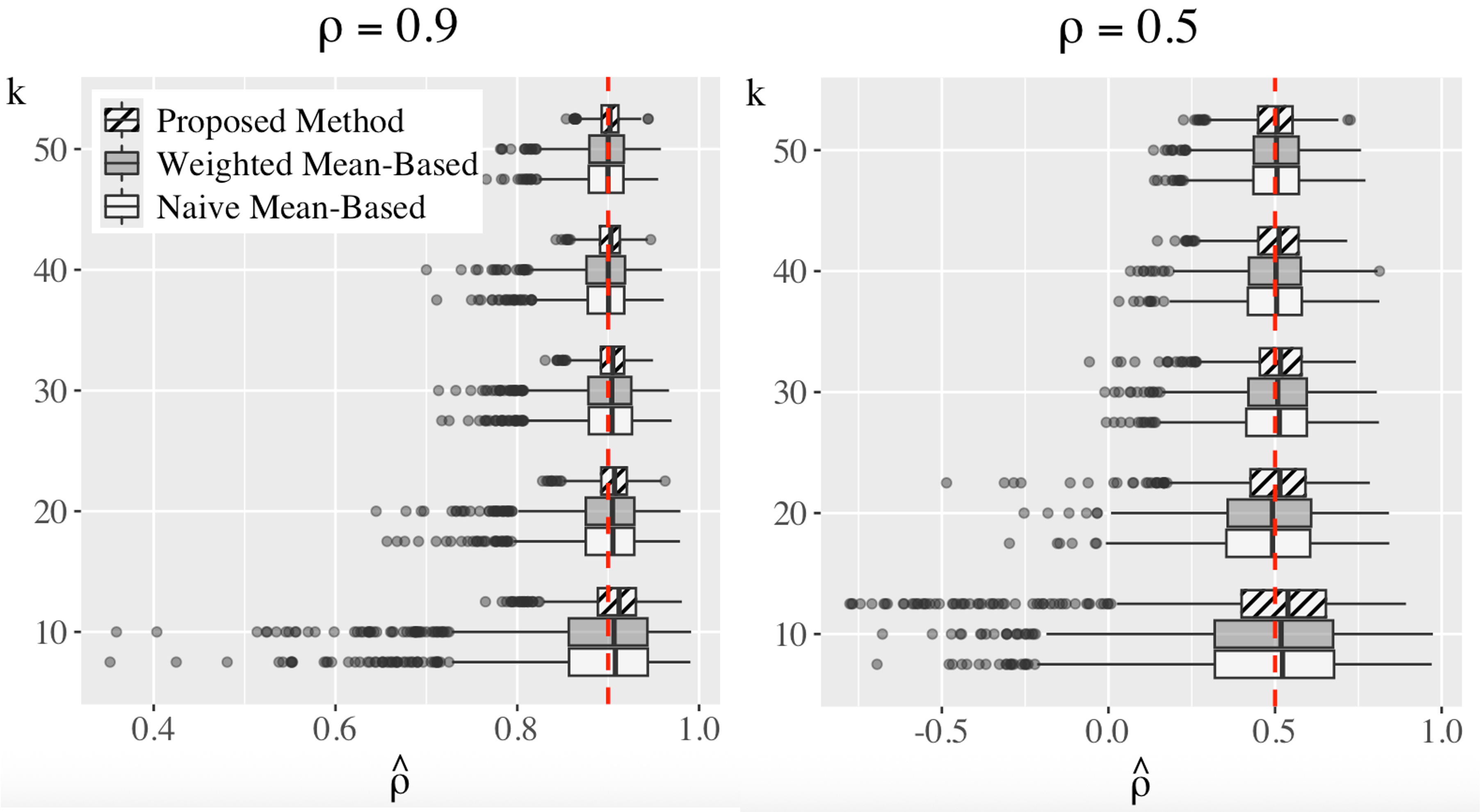}
    \caption{Boxplots of estimated correlation $\hat{\rho}$ under large-sample settings where $n_i \sim \mathrm{Uniform}(3000, 5000)$.}
    \label{fig:appendix-large-ni}
\end{figure}

\end{document}